\documentclass[shortAfour,times,sagev]{sagej}

\usepackage{moreverb,url}
\usepackage[utf8]{inputenc}

\usepackage[colorlinks,bookmarksopen,bookmarksnumbered,citecolor=red,urlcolor=red]{hyperref}

\usepackage{booktabs}
\usepackage{enumitem}
\usepackage{float}
\usepackage{epstopdf}

\newcommand\BibTeX{{\rmfamily B\kern-.05em \textsc{i\kern-.025em b}\kern-.08em
		T\kern-.1667em\lower.7ex\hbox{E}\kern-.125emX}}
	
\usepackage{wrapfig}
\usepackage{subcaption}
\usepackage{BayesMR} 

\def\journalname{Statistical Methods in Medical Research}
\def\volumeyear{2019}

\begin{document}
	
	\runninghead{Ioan Gabriel Bucur, Tom Claassen and Tom Heskes}
	
	\title{Inferring the Direction of a Causal Link and Estimating Its Effect via a Bayesian Mendelian Randomization Approach}
	
	\author{Ioan Gabriel Bucur\affilnum{1}, Tom Claassen\affilnum{1} and Tom Heskes\affilnum{1}}
	
	\affiliation{\affilnum{1}Data Science Department, Institute for Computing and Information Sciences, Radboud University, Nijmegen, The Netherlands\\}
	
	\corrauth{Ioan Gabriel Bucur, Faculty of Science, Radboud University, Postbus 9010, 6500 GL Nijmegen, The Netherlands.}
	
	\email{g.bucur@cs.ru.nl}

	\begin{abstract}
		The use of genetic variants as instrumental variables -- an approach known as Mendelian randomization -- is a popular epidemiological method for estimating the causal effect of an exposure (phenotype, biomarker, risk factor) on a disease or health-related outcome from observational data. Instrumental variables must satisfy strong, often untestable assumptions, which means that finding good genetic instruments among a large list of potential candidates is challenging. This difficulty is compounded by the fact that many genetic variants influence more than one phenotype through different causal pathways, a phenomenon called horizontal pleiotropy. This leads to errors not only in estimating the magnitude of the causal effect, but also in inferring the direction of the putative causal link. In this paper, we propose a Bayesian approach called \textsc{BayesMR} that is a generalization of the Mendelian randomization technique in which we allow for pleiotropic effects and, crucially, for the possibility of reverse causation. The output of the method is a posterior distribution over the target causal effect, which provides an immediate and easily interpretable measure of the uncertainty in the estimation. More importantly, we use Bayesian model averaging to determine how much more likely the inferred direction is relative to the reverse direction.
	\end{abstract}
	\maketitle

	\section{Introduction} \label{sec:introduction}
	
	Finding and estimating causal effects from various measured \textit{exposures} (phenotypes, biomarkers, risk factors) to disease and health-related \textit{outcomes} is a crucial task in epidemiology. If we knew there was a causal link from a modifiable exposure (e.g., the level of LDL-cholesterol) to an outcome of interest (e.g., the incidence of cardiovascular disease), we could intervene on the former to produce a desired effect on the latter. Moreover, if we knew the \textit{magnitude} of the causal effect, we could then predict the efficacy of treating the disease by altering the level of the exposure. 
	
	\textit{Randomized controlled trials} (RCTs) are the best means of establishing the causal relation between exposure and outcome.~\cite{fletcher_clinical_2012} In epidemiological research, RCTs have been used, for example, to identify causal risk factors or to determine the efficacy of a new drug. However, it is often not practical or ethical to perform an RCT, in which case we want to be able to extract causal information from observational epidemiological studies. Directly inferring causation from observational data is problematic since observed correlations (associations) do not imply any particular causal relationship on their own. The correlation between two variables $\exs$ and $\out$ can be explained by the existence of a causal link from $\exs$ and $\out$ (\textit{direct causation}) or from $\out$ to $\exs$ (\textit{reverse causation}), but it can also be explained by an  unobserved common cause (\textit{confounder}) influencing the two variables simultaneously. It can even be the result of a combination of causal and confounding effects.
	
	One method for finding and estimating the magnitude of causal effects from observational data that has recently gained traction in the field of epidemiology is \textit{Mendelian randomization} (MR). In MR, germline \textit{genetic variants} are used as proxy variables for environmentally modifiable exposures with the goal of making causal inferences about the outcomes of the modifiable exposure.~\cite{lawlor_mendelian_2008} The basic principle behind Mendelian randomization is that the differences between individuals resulting from genetic variation are not subject to the confounding or reverse causation biases that distort observational findings.~\cite{davey_smith_when_2017} The natural ``randomization'' of \textit{alleles} (variant forms) can thus be likened to the random allocation of treatment  performed in RCTs in the sense that groups defined by the value of a relevant genetic variant will experience an on-average difference in exposure, while not differing with respect to confounding factors.~\cite{davey_smith_mendelian_2014}
	
\begin{figure}[!htb]
		\begin{minipage}{0.49\linewidth}
			\centering
			\includegraphics[width=0.9\linewidth]{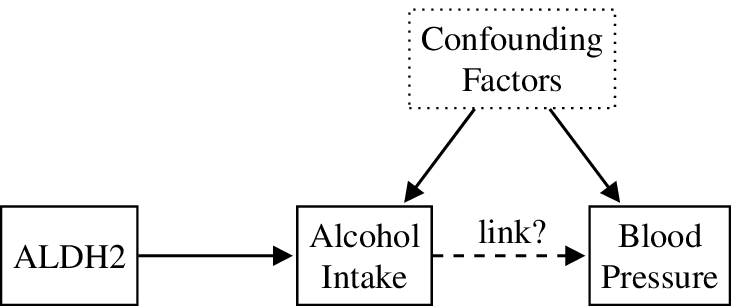}	
			\caption{Chen et al.~\cite{chen_alcohol_2008} used the ALDH2 genetic variant as a proxy for alcohol intake to determine if there is a causal link between the latter and blood pressure. They found that alcohol intake has a marked positive causal effect on blood pressure.}
			\label{fig:alcohol_bp}
		\end{minipage} \hfill
		\begin{minipage}{0.49\linewidth}
			\centering
			\includegraphics[width=0.9\linewidth]{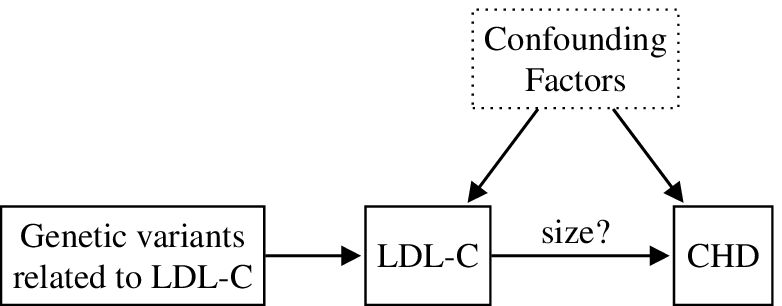}	
			\caption{Ference et al.~\cite{ference_effect_2012} used Mendelian randomization for estimating the (positive) causal effect of low-density lipoprotein cholesterol (LDL-C) on the risk of coronary heart disease (CHD). They estimated that for each $mmol/L$ lower LDL-C, the risk of CHD is reduced by 54.5\% (95\% CI [48.8\%, 59.5\%]).}
			\label{fig:chol_chd}
		\end{minipage}
	\end{figure}

	Mendelian randomization has already proved very useful in extracting causal information from observational studies. MR has been used successfully for drug target validation, for predicting side effects of drug targets, for discovering causal risk factors in clinical practice, or for a better understanding of complex molecular traits.~\cite{zheng_recent_2017} In a classic Mendelian randomization study, Chen et al.~\cite{chen_alcohol_2008} examined the nature of the association between alcohol intake and blood pressure. In this example, the exposure variable (alcohol intake) is a modifiable behavior, but conducting a randomized controlled trial would be problematic for both ethical and practical reasons: alcohol consumption is potentially damaging to the subject's health and measuring the intake is prone to error.  Chen et al.~\cite{chen_alcohol_2008} employed the genetic variant ALDH2 as a surrogate for measuring alcohol intake (see Figure~\ref{fig:alcohol_bp}). ALDH2 encodes \textit{aldehyde dehydrogenase}, a major enzyme involved in alcohol metabolism. The (*2*2) variant of the ALDH2 gene is associated with an accumulation of \textit{acetaldehyde} in the body after drinking alcohol, and therefore unpleasant symptoms. Because of this, carriers of this particular variant tend to limit their alcohol consumption. Chen et al.~\cite{chen_alcohol_2008} found that the individuals possessing the (*2*2) variant also have lower blood pressure on average. The authors considered the possibility that the (*2*2) \textit{genotype} influences blood pressure via a different causal pathway not mediated by alcohol intake (\textit{horizontal pleiotropy}). However, they found no association between ALDH2 and hypertension in females, for whom drinking levels in any genotype group were similar (very low). If there were pleiotropic effects, an effect on blood pressure would be seen in both sexes. The findings of Chen et al.~\cite{chen_alcohol_2008} support the idea that alcohol intake increases blood pressure and the risk of hypertension, to a much greater extent than previously thought. Their results indicate that previous observational epidemiological studies suggesting the cardioprotective effects of moderate alcohol consumption might have been misleading. 
	
	In a more recent study, Ference et al.~\cite{ference_effect_2012} performed an MR meta-analysis for the purpose of estimating the effect of long-term exposure to lower-plasma \textit{low-density lipoprotein cholesterol} (LDL-C) on the risk of \textit{coronary heart disease} (CHD). Unlike in the previous example, the causal link from LDL-C to CHD was already well established at the time of the study, but the magnitude of the (long-term) causal effect had not been reliably quantified (see Figure~\ref{fig:chol_chd}). To estimate the causal effect of LDL-C exposure on the risk of CHD, Ference et al.~\cite{ference_effect_2012} considered nine \textit{polymorphisms} (genetic variations resulting in the occurrence of several different alleles at a locus) in six different genes. For each polymorphism, they computed a causal effect estimate using Mendelian randomization and then they combined the results to obtain a more precise estimate. Finally, they compared the estimated causal effect of long-term exposure to lower LDL-C with the clinical benefit resulting from the same magnitude of LDL-C reduction during treatment with a statin. Ference et al.~\cite{ference_effect_2012} found that Mendelian random allocation to lower LDL-C levels was responsible for a 54.5\% reduction in CHD risk for each $mmol/L$ lower LDL-C. They concluded that  prolonged (lifetime) exposure to lower LDL-C results in a threefold greater reduction in the risk of CHD compared to the clinical benefit of a statin-based treatment producing the same magnitude of LDL-C reduction.
	
	\begin{wrapfigure}{r}{0.4\linewidth}
		\centering
		\includegraphics[width=0.9\linewidth]{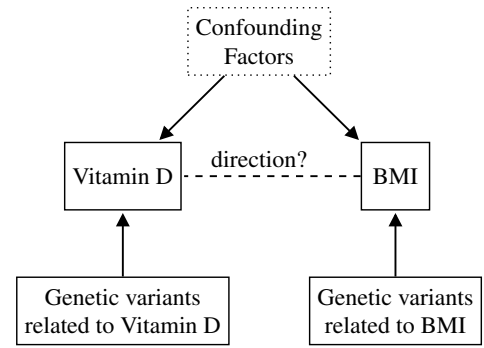}	
		\caption{Vimaleswaran et al.~\cite{vimaleswaran_causal_2013} explored the causal direction of the relationship between body mass index (BMI) and 25-hydroxivitamin D [25(OH)D] via bidirectional Mendelian randomization. They concluded that higher BMI leads to lower vitamin D levels and not the other way around.}
		\label{fig:bmi_vitamind}
	\end{wrapfigure}
	
	Finally, Vimaleswaran et al.~\cite{vimaleswaran_causal_2013} investigated the causality and direction of the association between body mass index (BMI) and 25-hydroxivitamin D [25(OH)D]. In this example, it is not clear whether the larger vitamin D storage capacity of obese individuals (vitamin D is stored in fatty tissues) lowers their 25(OH)D levels or whether 25(OH)D levels influence fat accumulation. The latter is an example of \textit{reverse causation}, which occurs when the outcome of interest precedes and leads to changes in the exposure instead of the other way around.~\cite{rothman_modern_2008} Reverse causation is an important source of bias in observational epidemiological studies,~\cite{davey_smith_mendelian_2003, lawlor_mendelian_2008} which can lead to misinterpretation in the observed association with respect to the potential impact of interventions.~\cite{wade_physical_2018} In their work, Vimaleswaran et al.~\cite{vimaleswaran_causal_2013} used a bidirectional Mendelian randomization approach (see Figure~\ref{fig:bmi_vitamind}), which implies performing an MR analysis in both directions, to distinguish between the two causal models. For this approach to work, it must be known on which phenotypic trait each genetic variant has a primary influence.~\cite{davey_smith_mendelian_2014} If vitamin D deficiency influences obesity, then the genetic variants primarily associated with lower 25(OH)D levels should also be associated with higher BMI. Conversely, if obesity leads to lower vitamin D levels, then the genetic variants primarily associated with higher BMI should also be related to lower 25(OH)D concentrations. Vimaleswaran et al.~\cite{vimaleswaran_causal_2013} concluded that higher BMI (obesity) leads to lower vitamin D levels and not the other way around. Their findings provided evidence for the role of obesity as a causal risk factor for vitamin D deficiency and suggested that interventions meant to reduce BMI are expected to decrease the prevalence of vitamin D deficiency, as later shown in intervention studies.~\cite{ibero-baraibar_increases_2015,gangloff_effect_2015}
	
	In this paper, we introduce a Bayesian framework (\textsc{BayesMR}) that extends the Mendelian randomization approach to situations where the direction of the causal effect between the two phenotypes of interest is unknown or uncertain. In our approach, we do not need to know in advance which of the two phenotypes is primarily influenced by the genetic variants. In the experimental section, we look at both directions of the association between smoking and coffee consumption using the same genetic variants in an attempt to determine which direction is more likely. This flexibility enables the researcher to apply the instrumental variable approach to a much broader set of genetic variants.  The rest of this paper is organized as follows. In Section~\ref{sec:assumptions}, we discuss the assumptions behind Mendelian randomization and recent extensions to the approach. Then, we introduce our proposed Bayesian model in Sections~\ref{sec:model} and~\ref{sec:selection}. We perform a series of simulations in Section~\ref{sec:simulation} to illustrate the strengths of our approach. In Section~\ref{sec:applications}, we apply our \textsc{BayesMR} method to a couple of real-world problems where the standard Mendelian randomization approaches might produce biased results. We conclude by discussing the advantages and disadvantages of our proposed method as well as potential applications in Section~\ref{sec:discussion}.

	\section{Instrumental variables and Mendelian randomization} \label{sec:assumptions}
	
	\subsection{Instrumental variable assumptions}
	
	\begin{wrapfigure}[10]{r}{0.4\linewidth}
		\centering
		\vspace{-4\baselineskip}
		\includegraphics[width=0.9\linewidth]{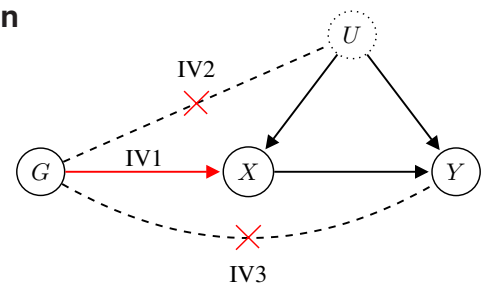}	
		\caption{Directed acyclic graph (DAG) showing the instrumental variable assumptions. Causal effects are unidirectional and denoted by arrows. The association entailed by \texttt{IV1} is highlighted in red, while the crossed-out dashed lines signify the absence of an association (causal or non-causal). The dotted border of $\conf$ means that the variable is unobserved. }
		\label{fig:iv_assumptions}
	\end{wrapfigure}
	
	Mendelian randomization is an approach in which genetic variants are used as instrumental variables for estimating the efffect of a modifiable phenotype or exposure on an outcome variable of clinical interest. A (valid) \textit{instrumental variable} (IV) or \textit{instrument}~\cite{bowden_instrumental_1990} is a factor which: 
	\begin{itemize}
		\item [(\texttt{IV1})] is associated with the exposure;
		\item [(\texttt{IV2})] is independent of all confounders of the exposure-outcome association;
		\item [(\texttt{IV3})] influences the outcome only through its effect on the exposure.
	\end{itemize}  
	
	The three core IV assumptions are depicted by the \textit{directed acyclic graph} (DAG) in Figure~\ref{fig:iv_assumptions}, where $\iv$ is the genetic variant used as an instrumental variable, $\exs$ is the variable measuring the exposure, $\out$ is the variable measuring the outcome of interest, and $\conf$ is an unmeasured variable that encapsulates the potential confounding effects affecting the exposure-outcome association. Note that for the graphical model to be acyclic, there must be no feedback loops in the system. Acyclicity is a standard assumption in Mendelian randomization and implies that the causal link of interest is unidirectional.~\cite{didelez_mendelian_2007} Assumption \texttt{IV1} is represented in Figure~\ref{fig:iv_assumptions} by the edge between $\iv$ and $\exs$. Note that the association between the genetic variant and the exposure need not be causal, although this is often assumed in the context of MR. Assumption \texttt{IV2} is denoted by a lack of paths from $\iv$ to $\conf$. Finally, assumption \texttt{IV3} is denoted by an absence of paths from $\iv$ to $\out$, except for the one that goes through $\exs$. Genetic variants are particularly suitable as candidate instrumental variables because they are fixed at conception and ``randomized'' in a population according to Mendel's laws of inheritance, meaning that their associations with non-genetic measures are generally not susceptible to confounding or reverse causation.~\cite{davey_smith_mendelian_2003, zheng_recent_2017} In graphical terms, this is equivalent to removing all arrows from non-genetic variables ($\exs, \out, \conf$ in Figure~\ref{fig:iv_assumptions}) to genetic variables ($\iv$ in  Figure~\ref{fig:iv_assumptions}).

	\subsection{The Mendelian randomization pipeline} \label{ssec:pipeline}
	
	The suitability of using genetic variants as instrumental variables combined with the wide availability of data on genetic associations extracted by means of high-throughput genomic technologies have contributed to the increasing popularity of Mendelian randomization applications.~\cite{davey_smith_mendelian_2014, visscher_10_2017} The first step in Mendelian randomization is to find genetic variants that are robustly associated with the exposure under study.~\cite{bowden_mendelian_2015} Ideally, MR is performed using a single variant whose biological effect on the exposure is well understood. In practice, however, such a single variant is not always available or known. Moreover, genetic variants typically only explain a small proportion of the variance in the exposure.~\cite{brion_calculating_2013} When a single reliable genetic instrument is not available, it has become common to use multiple genetic variants in MR studies. \textit{Genome-wide association studies} (GWAS) provide a rich source of potential instruments for MR analysis.~\cite{swerdlow_selecting_2016} Increasing the number of exposure-related genetic variants will increase the proportion of variance explained by the instruments, which will lead to improved precision in the causal estimate and to increased power in testing causal hypotheses.~\cite{bowden_mendelian_2015} What's more, the use of multiple genetic instruments provides opportunities to test for violations of the IV assumptions.~\cite{palmer_using_2012}
	
	Genetic variants are typically selected based on the size of the genetic effect on the exposure of interest and the specificity of the association.~\cite{swerdlow_selecting_2016} After deciding on a set of relevant exposure-related genetic instruments, the next step is to examine if they are also associated with the disease outcome. If no genotype-outcome association is found, then it becomes likely that there is no causal link from exposure to outcome. Otherwise, by making additional assumptions, it is possible to compute a consistent estimate of the causal relationship's strength.~\cite{didelez_mendelian_2007} A commonly made assumption is that the causal relation between the exposure $\exs$ and the outcome $\out$ is linear:
	\begin{equation} \label{eqn:linear_model}
	\out = \ce \exs + \err.
	\end{equation}
	The strength or magnitude of this relationship is then given by the parameter $\ce$, which is the \textit{causal effect} of the exposure on the outcome. In an MR analysis, the estimate of $\ce$ is derived from the associations with the genetic instruments. For example, when using a single genetic variant as an instrumental variable, a consistent estimate of the causal effect under the linear model assumption~\eqref{eqn:linear_model} is Wald's ratio~\cite{lawlor_mendelian_2008} (the Wald method~\cite{bowden_mendelian_2015}):
	\begin{equation} \label{eqn:wald_ratio}
	\hat{\ce}^{IV} = \frac{r_{\out\given\iv}}{r_{\exs\given\iv}},
	\end{equation}
	where $r_{\out\given\iv}$ is the coefficient of the regression of the outcome on the genetic variant and $r_{\exs\given\iv}$ is the coefficient of the regression of the exposure on the genetic variant. When using multiple genetic instruments, the individual IV estimates can be combined similarly to performing a meta-analysis:
	\begin{equation} \label{eqn:ivw}
	\hat{\ce}^{IVW} =  \frac{\sum_{j=1}^J \hat{\ce}^{IV}_j s_j^{-2}}{\sum_{j=1}^J s_j^{-2}},
	\end{equation}
	where $s_j$ is the standard error of $\hat{\ce}^{IV}_j$ and can be approximated using the delta method.~\cite{burgess_mendelian_2013} $\hat{\ce}^{IVW}$ is known as the \textit{inverse-variance weighted} (IVW) estimator and is equivalent to a zero-intercept weighted linear regression of the genetic associations with the outcome on the genetic associations with the exposure.~\cite{bowden_mendelian_2015} 
	
	Unfortunately, when selecting multiple genetic variants for Mendelian randomization based on their association with the exposure, we cannot be certain that all of them are valid instrumental variables. In the worst possible case, none of the selected variants are valid instruments for the exposure-outcome associations. As a result, there is great interest in developing methods robust to the validity of the IV assumptions.~\cite{davey_smith_mendelian_2014, zheng_recent_2017}

	\subsection{Challenges of Mendelian randomization}
	
	The application of MR methods requires genetic variants that are valid instruments for the exposure-outcome association, i.e., genetic variants satisfying the core IV assumptions. If a causal link from exposure to outcome exists, then a genetic variant associated with the exposure will also be associated with the outcome. Under the core IV assumptions, the converse also holds: an association between the genetic variant and the outcome implies the existence of a causal link from the exposure to the outcome.  When these assumptions are violated, however, we can no longer be sure that a causal link from exposure the outcome exists. A discovered genetic association with the disease outcome could be produced via a different causal pathway which is not mediated by the exposure (see Figure~\ref{fig:mr_pleiotropy}). In this case, the bias in the causal effect estimate obtained through the MR approach is potentially unbounded.~\cite{cornia_type-ii_2014} Even worse, if we are dealing with reverse causation for the exposure-outcome association, then the causal link whose effect we are trying to estimate does not exist (see Figure~\ref{fig:mr_reverse}).

	\begin{wrapfigure}{r}{0.35\linewidth}
		\centering
		\includegraphics[width=0.95\linewidth]{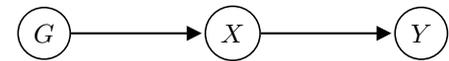}
		\caption{The Local Causal Discovery~\cite{cooper_simple_1997} (LCD) pattern can be discovered from observational data by testing if the genetic variant $\iv$ is independent from the outcome $\out$ when conditioning on the exposure $\exs$.} \label{fig:lcd}
	\end{wrapfigure}
	
	Unfortunately, only assumption \texttt{IV1} is directly testable from data. It is typically fulfilled by selecting only those genetic variants that are associated with an exposure of interest. In general, the validity of assumptions \texttt{IV2} and \texttt{IV3} cannot be established from the data, since they involve the unobserved confounder $\conf$.~\cite{didelez_mendelian_2007} However, if the exposure-outcome association is unconfounded (see Figure~\ref{fig:lcd}), then we can check the validity of \texttt{IV2} and \texttt{IV3} by performing a conditional independence test. In this particular scenario, the genetic variant becomes independent of the disease outcome after conditioning on the exposure variable. This principle of testing the IV assumptions by looking for a conditional independence is used to orient the causal relationship between the two traits in mediation-based approaches~\cite{hemani_orienting_2017} such as the \textit{likelihood-based causality model selection}~\cite{schadt_integrative_2005} (LCMS), the regression-based \textit{causal inference test}~\cite{millstein_disentangling_2009} (CIT) or the \textsc{Findr}~\cite{wang_efficient_2017} algorithm. In the artificial intelligence literature, this particular model is called the \textit{Local Causal Discovery}~\cite{cooper_simple_1997} (LCD) pattern. The testability of the LCD pattern has been exploited, for example, in the \textsc{Trigger}~\cite{chen_harnessing_2007} algorithm for the purpose of learning the structure of gene regulatory networks.
	
	Assumption \texttt{IV2} is not completely testable since there can always be unmeasured confounding variables for which we cannot account, but we can at least filter out the genetic variants associated with any potential confounding traits that have been measured.~\cite{yin_merp_2015} Fortunately, it is not very likely that genetic variants are associated with confounders of the exposure-outcome association.~\cite{davey_smith_clustered_2007} The assumption that is most likely to be violated is \texttt{IV3}, which states that the instrumental variable is only related to the outcome via its effect on the risk factor.~\cite{lawlor_commentary_2016} Assumption \texttt{IV3} does not hold if there are alternative causal pathways between the genetic variant and the disease outcome that are not mediated by the exposure (see Figure~\ref{fig:mr_pleiotropy}). This often happens when a genetic variant influences more than one phenotypic traits, a property called \textit{pleiotropy}.~\cite{van_kippersluis_pleiotropy-robust_2017}
	
	As the GWAS sample sizes continue to grow, the number of candidate genetic variants to be used as potential instruments is also increasing. Unfortunately, more and more of the supplied candidates are likely to be invalid instruments. Hemani et al.~\cite{hemani_automating_2017} show that horizontal pleiotropy and reverse causation can easily lead to invalid instrument selection in MR and suggest that the presence of horizontal pleiotropy should be considered the rule rather than the exception. This severely limits the application of Mendelian randomization to cases where previous background knowledge can be used to validate the core assumptions. By relaxing (some of) these assumptions, however, the MR approach can be applied to a much broader range of genetic variants.

	\subsubsection{Pleiotropy}
	
	In the genetics of complex diseases it is common for genetic variants to affect multiple disease-related phenotypes.~\cite{sivakumaran_abundant_2011, solovieff_pleiotropy_2013} If such a \textit{pleiotropic genetic variant} affects the disease outcome through different causal pathways, then \texttt{IV2} or  \texttt{IV3} or both are potentially violated. This makes pleiotropy problematic for MR studies.~\cite{didelez_mendelian_2007, davey_smith_mendelian_2014} As a result, many recent extensions of the MR approach have been aimed at producing better causal estimates in the presence of pleiotropic effects.~\cite{bowden_mendelian_2015, bowden_consistent_2016,  hartwig_robust_2017, li_mendelian_2017, van_kippersluis_pleiotropy-robust_2017, burgess_improving_2017,  berzuini_bayesian_2018} Pleiotropy can be divided into horizontal (biological) pleiotropy and vertical (mediated) pleiotropy. \textit{Horizontal pleiotropy} occurs when the same genetic variant affects multiple trait via different biological pathways, whereas \textit{vertical pleiotropy} occurs when the same genetic variant is associated with multiple traits found on the same biological pathway. We will only focus on the former, since only horizontal pleiotropy can lead to violations of \texttt{IV3}, and refer to it hereafter as `pleiotropy', for brevity.~\cite{van_kippersluis_pleiotropy-robust_2017}
	
	Two main approaches have been used for robustifying Mendelian randomization against potentially pleiotropic genetic variants. The first approach is to replace some of the (untestable) IV assumptions, e.g., \texttt{IV2} or \texttt{IV3} or both, with a weaker set of alternative assumptions. The second approach is to assume that the IV assumptions hold for some, but not necessarily for all the genetic variants. In this case, the invalid genetic instruments are allowed to violate the IV assumptions in an arbitrary way.~\cite{burgess_interpreting_2017}
	
	\textit{MR-Egger regression}~\cite{bowden_mendelian_2015} is a pleiotropy-robust method that falls into the first category. MR-Egger is an extension to the IVW method introduced in Subsection~\ref{ssec:pipeline}, in which an intercept term is added to the weighted linear regression of the genetic associations with the outcome on the genetic associations with the exposure. While IVW is consistent if the pleiotropic effects average to zero across the genetic variants, this extension allows for consistency even when the pleiotropic effects do not average to zero. Both the IVW and MR-Egger methods work under the \texttt{InSIDE} assumption, which stipulates that the association with the exposure is independent from the direct (pleiotropic) effect on the outcome across different genetic variants.~\cite{burgess_interpreting_2017} Another robust alternative to the IVW estimator is the \textit{weighted median estimator}~\cite{bowden_consistent_2016} (WME), for which we compute a weighted median of the IV estimates instead of a weighted average. The WME is a consistent estimate of the true causal effect as long as at least 50\% of the genetic instruments are valid. In a similar vein, Kang et al.~\cite{kang_instrumental_2016} have introduced a LASSO-type procedure to identify the valid instruments from within a set of candidate genetic variants. Hartwig et al.~\cite{hartwig_robust_2017} have proposed instead to compute the weighted mode of the individual IV estimates to obtain the \textit{mode-based estimate} (MBE). The MBE can be a consistent estimate of the true causal effect even when fewer than 50\% of the genetic instruments are valid. However, the application of the method relies on the assumption that  the largest number of similar individual IV estimates comes from valid instruments, an assumption termed the \textit{Zero Modal Pleiotropy Assumption} (\texttt{ZEMPA}). Finally, Hemani et al.~\cite{hemani_automating_2017} suggest combining these methods (along with others) in a ``mixture of experts'' machine learning approach.~\cite{bishop_pattern_2006}
	
	A number of Bayesian modeling approaches for dealing with pleiotropic genetic variants have been proposed, in which all variants used as instruments are allowed to exhibit pleiotropic effects on the outcome. Li~\cite{li_mendelian_2017} introduced empirical Bayes hierarchical models for estimating the causal effect in MR studies where many instruments are invalid, while Berzuini et al.~\cite{berzuini_bayesian_2018} performed a fully Bayesian treatment of the Mendelian randomization problem. In both cases, the authors handled pleiotropy by putting shrinkage priors on the pleiotropic effects, thereby forcing them to take values close to zero unless there is strong posterior evidence for pleiotropy.~\cite{burgess_inferring_2018} Thompson et al.~\cite{thompson_mendelian_2017}, on the other hand, modeled pleiotropy explicitly and used Bayesian model averaging to provide estimates that allow for uncertainty regarding the nature of pleiotropy.

	\subsubsection{Reverse causation} \label{ssec:reverse_causation}
	
	A crucial aspect that is often overlooked in Mendelian randomization studies is the implicit assumption made regarding the direction of the causal link between the exposure and the outcome.  While it is sometimes possible to rule out reverse causation, in most situations there is not enough background knowledge available about the underlying causal mechanism. The observed relationship between low levels of LDL cholesterol and risk of cancer is an example where we cannot a-priori assume a causal direction. It may be that low levels of LDL cholesterol are causal for cancer, but it is also possible that the presence of the disease has a negative effect on LDL cholesterol.~\cite{zheng_recent_2017} In the case of complex systems such as gene regulatory networks it is also unclear in which direction the gene regulation occurs.~\cite{aten_using_2008}

	\begin{figure}[H]
		\begin{minipage}{0.49\linewidth}
			\centering
			\includegraphics[width=0.8\linewidth]{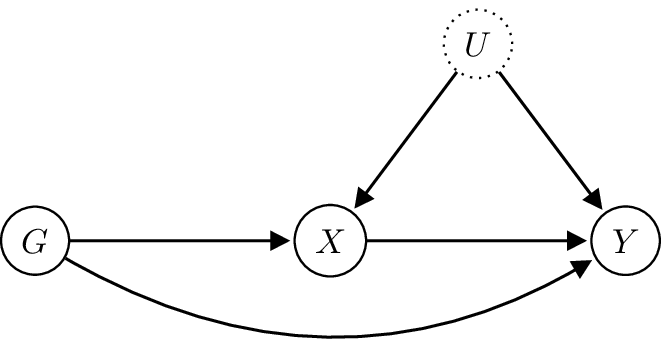}
			\caption{The genetic variant $\iv$ is pleiotropic since it affects both phenotypic measures ($\exs$ and $\out$). In this illustration, $\iv$ exhibits both \textit{vertical pleiotropy} ($\exs$ and $\out$ are influenced by $\iv$ via the same causal pathway $\iv \to \exs \to \out$) and \textit{horizontal pleiotropy} ($\exs$ and $\out$ are influenced by $\iv$ via different causal pathways). The former is crucial to the application of MR, while the latter results here in a violation of the \texttt{IV3} assumption. \vspace{\baselineskip} }
			\label{fig:mr_pleiotropy}
		\end{minipage} \hfill
		\begin{minipage}{0.49\linewidth}
			\centering
			\includegraphics[width=0.8\linewidth]{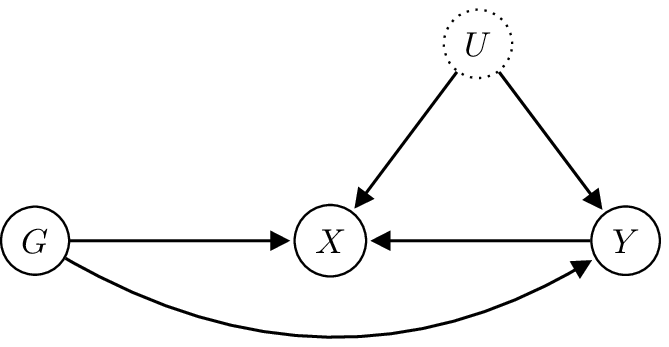}
			\caption{Reverse causation refers to the situation when the outcome $\out$ precedes and causes the exposure $\exs$ instead of the other way around.~\cite{flegal_reverse_2011} The term `reverse' refers to the fact that the direction of the causal effect is opposite to what was expected based on the study design. We emphasize that although we denote the exposure by $\exs$ and the outcome by $\out$, these terms do not have any causal meaning, i.e., they do not imply a particular causal ordering}.
			\label{fig:mr_reverse}
		\end{minipage}
	\end{figure}
	
	The existence of alternative causal paths from a genetic variant to the outcome of interest affects not only the estimation of causal effects, but also model identification. If there exists a causal pathway from the genetic variant to the outcome that is not mediated by the exposure, then an exposure-outcome causal link is no longer required to explain their association (see Figure~\ref{fig:mr_pleiotropy}). In fact, the causal link between the two traits may have the opposite direction (see Figure~\ref{fig:mr_reverse}). Ignoring the possibility of reverse causation for the exposure-outcome association is potentially dangerous. If the true causal link is in the expected direction (from $\exs$ to $\out$, Figure~\ref{fig:mr_pleiotropy}), we will be underestimating the uncertainty of our estimate by not considering the alternative model. If the true causal link is in the opposite direction (from $\out$ to $\exs$, Figure~\ref{fig:mr_reverse}), then the MR approach will produce a misleading result, as we are trying to estimate the wrong causal effect. Bidirectional Mendelian randomization~\cite{davey_smith_mendelian_2014} is one approach used to distinguish between an ``exposure'' having a causal effect on an ``outcome'' and the reverse. The idea is to perform MR in both directions to tease apart the causal relationships $\exs \rightarrow \out$ and $\out \rightarrow \exs$.~\cite{zheng_recent_2017} To make use of this approach, it is essential to know if a genetic variant primarily influences the ``exposure'' or the ``outcome''. However, this can be difficult to assess when utilizing variants with little or no understanding of their biological effects.~\cite{davey_smith_mendelian_2014} 
	
	Agakov et al.~\cite{agakov_sparse_2010} proposed a Bayesian framework built on sparse linear methods developed for regression, which is similar in spirit to \textsc{BayesMR}. In their approach, they first produce a sparse representation of the data by putting a Laplace prior on the linear coefficients and performing MAP inference. They then compute the evidences (marginal likelihoods) using the Laplace approximation for the direct, reverse, and pleiotropic models, with and without latent confounders.  This \textit{sparse instrumental variables} (SPIV) approach has been used to infer the direction of causality between vitamin D levels (plasma 25-hydroxyvitamin D) and colorectal cancer.~\cite{zgaga_model_2013} More recently, Hemani et al.~\cite{hemani_orienting_2017} have suggested using the \textit{Steiger Z-test}~\cite{steiger_tests_1980} of correlated correlations for elucidating the direction of the causal link. Their \textit{MR Steiger}~\cite{hemani_orienting_2017} approach is based on the simple principle that in most circumstances the genetic variant will exhibit stronger correlation with the trait located upstream in the causal chain.

	\subsection{Summary}
	
	Mendelian randomization is a powerful principle that can help strengthen causal inference by combining observational data with genetic knowledge. However, MR methods require good candidate genetic variants to be used as instruments, which often limits their application to variants with well understood biology. The wide spread of pleiotropic genetic variants, which violate key IV assumptions (\texttt{IV2} and especially \texttt{IV3}), is a particularly troublesome issue.~\cite{davey_smith_mendelian_2014, hemani_evaluating_2018} Many methods have been designed to improve causal effect estimation in the presence of pleiotropic effects,~\cite{li_mendelian_2017, thompson_mendelian_2017, berzuini_bayesian_2018} but they do not incorporate the possibility of reverse causation. At the same time, a number of methods have been developed for inferring the correct direction of the causal link between two phenotypes,~\cite{hemani_automating_2017, hemani_orienting_2017} but these do not provide a measure of uncertainty in the model selection. In the following section, we introduce a general method with which we aim to handle all of these issues simultaneously. We first perform Bayesian model averaging to account for the uncertainty in the direction of the causal link and then, for each model, we obtain an informative posterior distribution for the causal effect of interest.

	\section{Bayesian model specification} \label{sec:model}
	
	\subsection{Setup}
	
	\begin{wrapfigure}[15]{r}{0.5\linewidth}
		\centering
		\vspace{-4\baselineskip}
		\includegraphics[width=\linewidth]{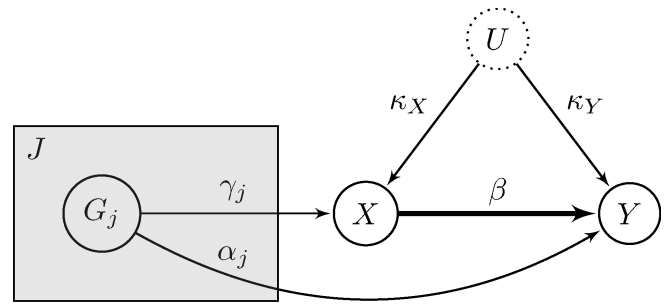}
		\caption{Graphical description of our assumed generative model. We denote the exposure variable by $\exs$ and the outcome variable by $\out$. We are interested in the causal effect from $\exs$ to $\out$, which is denoted by $\ce$. The association between $\exs$ and $\out$ is obfuscated by the unobserved variable $\conf$, which we use to model unmeasured confounding explicitly. The shaded plate indicates replication across the $J$ independent genetic variants $\iv_j, j \in \{1, 2, ..., J\}$. Note that the replication also applies to the parameters $\is_j$ and $\ply_j$ and their corresponding edges.}  \label{fig:model}
	\end{wrapfigure}

	We assume that the data is generated from the model illustrated in the causal diagram from Figure~\ref{fig:model}, where all relationships between variables are linear with no effect modification. We consider $J$ independent genetic variables, which we denote by $\iv_j$,  $j \in \{1, 2, ..., J\}$. Their genetic variation is given by the \textit{allele} (variant form) found at a specific locus. The two allele copies inherited at a specific locus form the \textit{genotype} of an individual. Here we only consider binary \textit{polymorphisms}, i.e., each allele has two possible forms.~\cite{lawlor_mendelian_2008} For convenience will refer to these forms as \textit{reference alleles} and \textit{effect alleles}, without assigning any biological meaning to the terms. Each variable $\iv_j$ can take the values 0, 1 or 2 corresponding to the number of effect alleles, as opposed to reference alleles, at the locus. Their corresponding \textit{effect allele frequencies} (EAF) are denoted by $\eaf_j$, $j \in \{1, 2, ..., J\}$. 
	
	The generative model from Figure~\ref{fig:model} is described by the following linear structural equation system:
	\begin{equation} \label{eqn:sem}
	\begin{aligned}
	\exs &:= \mu_\exs + \isv^\trp \ivv + \cc_\exs \conf +  \err_\exs \\ 
	\out &:= \mu_\out + \plyv^\trp \ivv + \ce \exs + \cc_\out \conf + \err_\out \\ 
	\end{aligned},
	\end{equation} where $\ivv$ is the random vector containing the $J$ independent variables $\iv_j$, $\isv$ is the corresponding vector of instrument strengths, while the vector $\plyv$ collects the potential pleiotropic effects of the genetic variants. $\ce$ represents the putative causal effect of the exposure $\exs$ on the outcome $\out$, which is what we wish to estimate. $\cc_\exs$ and $\cc_\out$ represent the effect of the confounder $\conf$ on $\exs$ and $\out$, respectively. Finally, $\mu_\exs$ and $\mu_\out$ are the average values of the exposure and outcome variables, while $\err_\exs$ and $\err_\out$ represent independent measurement error. 
	
	In this work, we restrict our attention to the case where the exposure and outcome variables are continuous. We also assume without loss of generality that $\mu_\exs = \mu_\out = 0$. We denote the variance of the intrinsic error terms by $\Var{\err_\exs} = \sigma_\exs^2$ and $\Var{\err_\out} = \sigma_\out^2$. Without loss of generality, we can fix the mean of the confounding variable and of the error terms to zero, i.e., $\EV{\conf} = \EV{\err_\exs} = \EV{\err_\out} = 0$. Also w.l.o.g., we can fix the variance of the confounding variable to one ($\Var{\conf} = 1$) by appropriately rescaling the confounding coefficients $\cc_\exs$ and $\cc_\out$.

	\subsection{Model assumptions}
	
	The first two core IV assumptions are implied by our proposed model. The existence of direct edges from each $\iv_j$ to the exposure $\exs$ implies that \texttt{IV1} is satisfied for each genetic variant. At the same time, we assume that for any $j \in \natseq{J}$ we have $\iv_j \indep \conf$, where we use the notation $X \indep Y \given \bfS$ \cite{dawid_conditional_1979} to denote that $X$ is conditionally independent of $Y$ given $\bfS$. These independence statements signify that the genetic variants are not correlated with any confounding variables (\texttt{IV2}). On the other hand, we allow for unrestricted violations of assumption \texttt{IV3}, as long as they do not lead to violation of \texttt{IV2}. We summarize the contribution of the pleiotropic effects in our system with a single directed edge between each genetic variant $\iv_j$ and the outcome variable $\out$ (See Figure~\ref{fig:model}). The strength of the pleiotropic effect is denoted by $\ply_j$. A genetic variant $\iv_j$ is a valid instrumental variable only if its direct effect on the outcome is exactly zero, i.e., $\alpha_j = 0$. 
	
	In our model, we assume that the instrument strengths $\is_j$ and the pleiotropic effects $\ply_j$ are independent a-priori. Berzuini et al. have termed this \textit{Instrument Effects Orthogonality} (\texttt{IEO}),~\cite{berzuini_bayesian_2018} which is the Bayesian interpretation of the \textit{Instrument Strength Independent of Direct Effect} (\texttt{InSIDE}) assumption.~\cite{bowden_mendelian_2015} This relaxation of the stronger \texttt{IV3} assumption is frequently used in pleiotropy-robust MR methods~\cite{zheng_recent_2017, verweij_investigating_2018} and is in line with the principle of independent (causal) mechanisms.~\cite{peters_elements_2017} Here, we distinguish between the causal mechanism describing the direct association between the genetic variants and $\exs$, which is parametrized by $\is_j$, and the causal mechanism describing the direct (not through $\exs$) association between the genetic variants and $\out$, which is parametrized by $\ply_j$.
	
	 We also assume that the genetic variants are independent, i.e., $\iv_i \indep \iv_j, \forall i, j \in \natseq{J}, i \neq j$, although our model could be easily extended to handle correlated genetic variants by introducing additional parameters. This means that we do not allow for the possibility of \textit{linkage disequilibrium}, where one genetic variant can potentially have a causal effect on another related genetic variant. We do not assume, however, that the genetic variants primarily influence one phenotype or the other, as required for instance in the application of standard bidirectional Mendelian randomization. This enables us to use the same set of instruments and the same prior distributions for both the forward and reverse analyses, as the variables $\exs$ and $\out$ become essentially interchangeable.

	\subsection{Likelihood}
	
	We assume that the data generating process is described by equation system~\eqref{eqn:sem} and that each data sample is a complete set of observed values for the exposure $\exs$, the outcome $\out$ and the genetic variants $\ivv \equiv [\iv_1, \iv_2, ... \iv_j]^\trp$. The input data set $\data$ consists of $N$ i.i.d. samples of $(\ivv, \exs, \out)$. Our model parameters consist of: 
	\begin{itemize}
		\item $(\isv, \ce, \plyv)$ -- the structural coefficients denoting the direct causal effects from $\ivv$ to $\exs$ (instrument strengths), from $\exs$ to $\out$ (the target causal effect), and from $\ivv$ to $\out$ (pleiotropic effects), respectively;
		\item $(\sigma_\exs, \sigma_\out)$ -- the scale parameters of the intrinsic error terms $\err_\exs$ and $\err_\out$;
		\item $(\cc_\exs, \cc_\out)$ -- the coefficients corresponding to the confounding effect of $\conf$.
	\end{itemize} %
	We collect all parameters of the model in the vector $\parm = (\isv, \ce, \plyv, \sigma_\exs, \sigma_\out, \cc_\exs, \cc_\out)$. 
	
	We model each genetic variable $\iv_j$ as binomially distributed, where we count the number of effect alleles at the genetic locus, either 0, 1 or 2. If $\eaf_j$ is the EAF of $\iv_j$, then the binomial distribution has parameters $n = 2$ and $p = \eaf_j$, so $\Var{G_j} = 2 \cdot \eaf_j \cdot (1 - \eaf_j)$. In our model, the intrinsic error terms of $\exs$ and $\out$ are independent and follow a Gaussian distribution: $\err_\exs \sim \normd(0, \sigma_\exs^2), \err_\out \sim \normd(0, \sigma_\out^2)$. The confounding variable $\conf$ follows a standard normal distribution: $\conf \sim \normd(0, 1)$. Note that we can set the variance of $\conf$ to one without any loss of generality by appropriately rescaling the confounding coefficients $\cc_\exs$ and $\cc_\out$. We obtain a linear-Gaussian model conditional on the value of the genetic variants, in the same vein as Jones et al.,~\cite{jones_choice_2012} Li,~\cite{li_mendelian_2017} or Berzuini et al.:~\cite{berzuini_bayesian_2018}
	\begin{equation}
	\left. \begin{bmatrix} \exs \\ \out \end{bmatrix} \right| \ivv \sim \mathcal{N}(\boldsymbol{\mu}, \mathbf{\Sigma}), \textrm{ where } \bmu = \begin{bmatrix} \isv^\trp \ivv \\ (\plyv + \isv \ce)^\trp \ivv \end{bmatrix},  \bSigma = \scriptstyle \begin{bmatrix} \sigma^2_\exs + \cc_\exs^2 & \cc_\exs \cc_\out + \ce (\sigma_\exs^2 + \cc_\exs^2) \\ \cc_\exs \cc_\out + \ce (\sigma_\exs^2 + \cc_\exs^2) & \sigma^2_\out + \ce^2 \sigma_\exs^2 + (\cc_\out + \ce \cc_\exs)^2 \\  \end{bmatrix}.
	\end{equation}
	
	Given the conditional linear-Gaussian model, the log-likelihood is equal, up to constant terms, to:
	\begin{equation}
	\log\lik_\data \propto -\frac{N}{2} (\log\det{\bSigma} + \textrm{tr}\{\bSigma^{-1} \bfS\}),
	\end{equation}
	where $\bfS = \dfrac{1}{N} \sum\limits_{i=1}^N \left(
	\begin{bmatrix} \exs_i - \isv^\trp \ivv_i \\ \out_i - (\plyv + \isv \beta)^\trp \ivv_i \end{bmatrix} \begin{bmatrix} \exs_i - \isv^\trp \ivv_i & \out_i - (\plyv + \isv \beta)^\trp \ivv_i \end{bmatrix}
	\right)$. We use $\ivv_i, \exs_i, \out_i$ to denote the $i$-th observed value of the genetic vector, exposure variable, and outcome variable, respectively.
	
	Both Berzuini et al.~\cite{berzuini_bayesian_2018} and Li~\cite{li_mendelian_2017} consider the same linear-Gaussian model, but use a slightly different parametrization: $\parm' = (\isv, \ce, \plyv, \cc_\exs^2 + \sigma_\exs^2, \cc_\out^2 + \sigma_\out^2, \cc_\exs \cc_\out)$. The difference between our approach and theirs is that we model the confounding explicitly by defining the linear confounding effects $\cc_\exs$ and $\cc_\out$, while they model the confounding implicitly by defining a nonzero covariance parameter for $(\exs, \out - \beta \exs) \given \ivv$. In other words, using the terminology from Jones et al.~\cite{jones_choice_2012}, we employ a ``full model'' parametrization, while Li~\cite{li_mendelian_2017} and Berzuini et al.~\cite{berzuini_bayesian_2018} consider a parametrization equivalent to the ``correlated errors model''. In both the full and correlated errors model, the likelihood of the observables $(\exs, \out)$ given $\ivv$ is bivariate normal.~\cite{jones_choice_2012} Thompson et al.~\cite{thompson_mendelian_2017} also assume a linear-Gaussian model, but approximate the discrete distribution over the set of genetic variants with a continuous normal distribution.

	\subsection{Priors} \label{ssec:prior}
	
	According to Jones et al.,~\cite{jones_choice_2012} informative priors are crucial to the success of Bayesian MR methods, regardless of whether or not they are identifiable, and so-called `vague' (uninformative) priors should be avoided whenever possible. Thompson et al.~\cite{thompson_mendelian_2017} mention two arguments against the use of an uninformative prior in the Mendelian randomization context: 1. that the analysis is already subjective due to the selection of the genetic variants and; 2. the data alone is sufficient to produce useful estimates of the causal effect. However, care must be taken when specifying informative priors, as substantial variation in parameter estimates can occur as the prescribed priors are modified. 
	
	For most genetic variants, it is unrealistic to expect exactly zero pleiotropy, i.e., that the effect on the outcome variable is completely mediated by the exposure.~\cite{verbanck_detection_2018} However, we can differentiate between pleiotropic effects that are weak and strong relative to the other interactions. A weak pleiotropic effect will not introduce significant bias, and therefore still enable us to produce a useful estimate of the causal effect, in spite of the violation of \texttt{IV3}.~\cite{zhu_causal_2018} For this reason, in choosing the prior for our structural parameters, we start from the assumption that the interactions between variables are either `strong' (relevant) or `weak' (irrelevant), in a similar vein to the considerations made by Berzuini et al.~\cite{berzuini_bayesian_2018} This is different from the standard assumption that the pleiotropic effects are either zero or nonzero, as used for example in the LASSO-based method of Kang et al.~\cite{kang_instrumental_2016} A reasonable choice for the prior (of $\pare \in \parm$) is then a scale mixture of two Gaussian distributions centered at zero:
	\begin{equation} \label{eqn:gauss_mix_prior}
	\pare \sim \wt \cdot \normd(0, \tau^2) + (1 - \wt) \cdot \normd(0, \lambda \tau^2), 
	\end{equation} where $\tau > 0$ and $\lambda \in (0, 1)$. In the limit of $\lambda \rightarrow 0$, we obtain the spike-and-slab prior.~\cite{ishwaran_spike_2005} This Gaussian mixture prior has a straightforward interpretation, in which the component with higher variance is meant to capture the `strong' interactions, while the component with lower variance is meant to capture the `weak' interactions. It is informative in the sense that we expect some of the interactions in our data to be negligible, though not exactly zero, and therefore encourage the exclusion of irrelevant nonzero effects.~\cite{george_variable_1993} For example, we might expect that some (most) of the genetic variants do not have significant pleiotropic effects, but exhibit some weak unmediated residual interaction.
	
	Our Gaussian mixture prior induces model parsimony by giving a preference for the model with the smallest number of strong parameters, which will then allow us to choose the most parsimonious causal explanation (direct causation, reverse causation, no causation).~\cite{bucur_robust_2017} Moreover, since every interaction is shrunk selectively, in the sense that smaller coefficient estimates are shrunk towards zero more sharply than larger coefficients, a significant gain in efficiency can be achieved.~\cite{zhao_powerful_2018} Other shrinkage priors, such as the horseshoe employed by Berzuini et al.~\cite{berzuini_bayesian_2018}, have also been used successfully for Bayesian MR. However, we have chosen the spike-and-slab prior due to ease of implementation and interpretation. In our experiments, we found that nested sampling works better when we specify the structural parameter (marginal) prior density directly instead of using a hierarchical approach. Consequently, we found it helpful that the spike-and-slab prior density is simply a mixture of Gaussian densities, as opposed to the horseshoe prior density which has no closed-form expression.~\cite{carvalho_handling_2009}

	\subsubsection{Rescaling}
	
	To make our method independent of the measured variables' scale, we rescale our model parameters as follows. We divide the first equation in~\eqref{eqn:sem} by $\sigma_\exs$, the scale of the first error term $\err_\exs$, and we divide the second equation by $\sigma_\out$, the scale of the second error term $\err_\out$:
	
	\begin{equation} \label{eqn:ssem}
	\begin{aligned}
	\rsc{\exs} &:= \rsc{\mu}_\exs + \rsc{\is}^\trp \rsc{\ivv} + \rsc{\cc}_\exs \conf +  \rsc{\err}_\exs \\ 
	\rsc{\out} &:= \rsc{\mu}_\out + \rsc{\plyv}^\trp \rsc{\ivv} + \rsc{\ce} \rsc{\exs} + \rsc{\cc}_\out \conf + \rsc{\err}_\out \\ 
	\end{aligned}.
	\end{equation}
	The standardized variables %
	$$
	\rsc{\exs} = \frac{\exs}{\sigma_\exs}, \quad 
	\rsc{\out} = \frac{\out}{\sigma_\out}, \quad 
	\rsc{\iv}_j = \frac{\iv_j}{\sqrt{2 \eaf_j (1 - \eaf_j)}}, \forall j \in \natseq{J},
	$$ %
	are dimensionless quantities. The parameters of the model are also appropriately rescaled and become dimensionless:
	
	$$
	\rsc{\ce} = \frac{\sigma_\exs}{\sigma_\out} \ce, \quad
	\rsc{\is}_j = \frac{\sqrt{2 \eaf_j (1 - \eaf_j)}}{\sigma_\exs} \is_j, \quad
	\rsc{\ply}_j = \frac{\sqrt{2 \eaf_j (1 - \eaf_j)}}{\sigma_\out} \is_j.
	$$
	We then put priors on the set of rescaled parameters $\sparm = (\rsc{\isv}, \rsc{\ce}, \rsc{\plyv}, \sigma_\exs, \sigma_\out, \rsc{\cc}_\exs, \rsc{\cc}_\out)$.

	\subsubsection{Choice of priors}
	
	\begin{itemize}
		\item[$\rsc{\plyv}:$] A Gaussian mixture prior suitably reflects the uncertainty in the pleiotropy of a genetic variant. If the pleiotropic effect is weak, then the induced bias for computing the causal effect is very small, even though the \textit{exclusion restriction} (\texttt{IV3}) assumption is violated. On the other hand, if the pleiotropic effect is strong, we would like to correct for this bias. Our model incorporates both situations naturally by assigning a Gaussian mixture (spike-and-slab) prior to the (rescaled) pleiotropic interaction. If the interaction is weak (irrelevant), then it will be captured by the `spike'. If the interaction is strong (relevant), then it will be captured by the `slab'.
		
		\item[$\rsc{\isv}:$] The genetic variants are often preselected using the criterion of association with the exposure of interest, which suggests that the instrument strength is nonzero. A normal prior with large enough variance would be appropriate for this parameter. If we have prior knowledge that the instrument is weak, we can appropriately choose a smaller variance for the prior. It is also possible that variants selected for the association with the exposure are actually more strongly related (upstream) to the outcome. The strength of the genotype-exposure association could then be seen as `weak' or `irrelevant' compared to the genotype-outcome association. With genome-wide association studies growing ever larger, the statistical power to detect significant associations that may be influencing the trait downstream of many other pathways increases,~\cite{hemani_automating_2017} in which case a Gaussian mixture prior could also be appropriate.
		
		\item[$\rsc{\ce}:$] The mixture of Gaussians prior is also suitable for the exposure-outcome causal effect, since it is possible that the dependence $\iv \nindep \out$ is explained by pleiotropy and not by a causal effect $\exs \rightarrow \out$. Because we do not know a-priori if we should expect a significant causal effect, we fix the weights of the mixture to $0.5$, which corresponds to an indifference prior. \cite{ishwaran_spike_2005}
		
		\item[$\rsc{\cc}_\exs, \rsc{\cc}_\out:$] We also put a Gaussian mixture prior on the confounding coefficients to express the uncertainty with regard to the presence of confounding.
		
		\item[$\sigma_\exs, \sigma_\out:$] For the scale parameters, we choose a weakly informative, but proper half-Gaussian prior with a large standard deviation, as suggested by Gelman.~\cite{gelman_prior_2006}
	\end{itemize}

	\subsubsection{Prior hyperparameters}
	
	When we do not know if the (potentially pleiotropic) genetic variants we intend to use for Mendelian randomization are valid or not, it is unclear whether their pleiotropic effects are more likely to be `large' or `small'. To express this uncertainty, we define the hyperparameter $\wt_{\plyv}$, corresponding to the weight of the Gaussian mixture prior for the pleiotropic effects and assign to it an uninformative uniform hyperprior:
	\begin{align*}
	\tilde{\ply}_j &\sim \wt_{\plyv} \cdot \normd(0, \tau^2) + (1 - \wt_{\plyv}) \cdot \normd(0, \lambda \tau^2), \; \forall j \\ 
	\wt_{\plyv} & \sim \unifd(0, 1).
	\end{align*}
	We can make a similar argument for the strength of the association between the genetic variants and the exposures, i.e., the `instrument strengths'. Analogously, we obtain the hierarchical prior:
	\begin{align*}
	\tilde{\is}_j &\sim \wt_{\isv} \cdot \normd(0, \tau^2) + (1 - \wt_{\isv}) \cdot \normd(0, \lambda \tau^2), \; \forall j \\ 
	\wt_{\isv} & \sim \unifd(0, 1).
	\end{align*}

	\section{Bayesian model averaging and inference} \label{sec:selection}
	
	\subsection{Computing the model evidence}
	
	The Bayesian view of model comparison involves the use of probabilities to represent uncertainty in the choice of model.~\cite{bishop_pattern_2006} We define the prior probability of a model $p(\model)$ to express our preference for different models. We then wish to compute the posterior distribution $$ p(\model \given \data) \propto p(\data \given \model) p(\model),$$ which will give us the probability of the model given the data. Bayesian model comparison and selection is a challenging problem, as it involves computing the Bayesian \textit{model evidence}, also called the \textit{marginal likelihood}:~\cite{bishop_pattern_2006}
	\begin{equation}
	p(\data \given \model) = \int \diff \parm p(\data \given \model, \parm) p(\parm \given \model). \label{eqn:marglik}
	\end{equation} 
	
	The model evidence can be viewed as the probability of generating the data set $\data$ from a model whose parameters are sampled at random from the prior, and therefore expresses the preference shown by the data for different models. Evaluating the model evidence involves marginalizing the likelihood-prior product over all the parameters of the model (see Equation~\eqref{eqn:marglik}). With the exception of a few special cases, this marginalization is analytically intractable and can potentially become very difficult to compute as the number of parameters increases. Fortunately, there are sampling schemes such as \textit{nested sampling}~\cite{skilling_nested_2006} that are designed to accurately compute the model evidence. 
	
	When performing model comparison, it is convenient to work with probability ratios. The \textit{prior odds ratio} $\frac{p(\model)}{p(\model')} $ expresses the preference we give to model $\model$ over model $\model'$ before looking at any of the data. The ratio of evidences for the two competing models, $\frac{p(\data \given \model)}{p(\data \given \model')}$, is called the \textit{Bayes factor}.~\cite{bishop_pattern_2006} Combining the prior odds with the Bayes factor, we get the \textit{posterior odds ratio}: $$ \frac{p(\model \given \data)}{p(\model' \given \data)} = \frac{p(\data \given \model)}{p(\data \given \model')} \cdot \frac{p(\model)}{p(\model')}.$$
	
\begin{figure}[H]
		\centering
		
		\begin{subfigure}{0.49\linewidth}
			\centering
			\includegraphics[width=\linewidth]{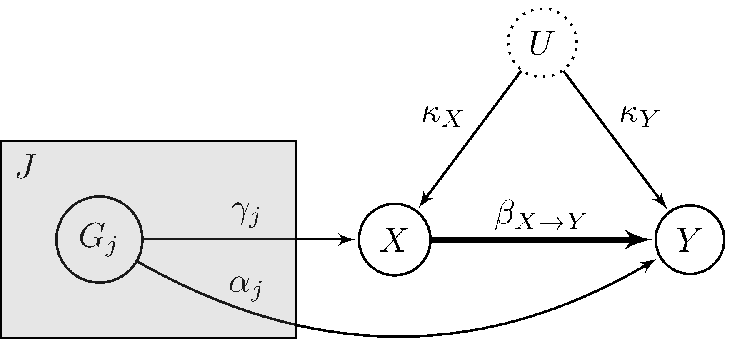}
			\caption{$(\model_\dir)$ Model where the causal relationship is in the expected direction. $\ce_\dir$ is the parameter measuring the (linear) causal effect of the exposure $\exs$ on the outcome $\out$. The shaded plate indicates replication across the variants $\iv_j$.} \label{subfig:expected_direction}
		\end{subfigure} \hfill
		\begin{subfigure}{0.49\linewidth}
			\centering
			\includegraphics[width=\linewidth]{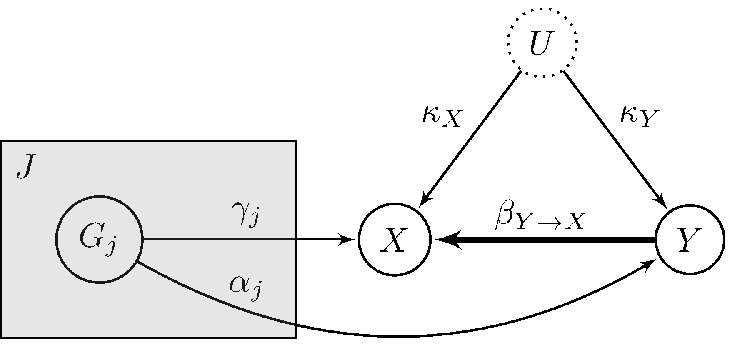}
			\caption{$(\model_\rev)$ Alternative model, where the causal relationship is in the reverse direction, from outcome to exposure. Note that $\ce_\rev$ is a different parameter than $\ce_\dir$. The latter is equal to zero for this model, as there is no causal link from $\exs$ to $\out$.} \label{subfig:reverse_direction}
		\end{subfigure}
		\caption{Competing models encompassing both possible directions for the causal link between exposure and outcome.} \label{fig:competing_models}
	\end{figure}

	In our approach, we compare two competing models corresponding to the two possible causal orderings for the non-genetic variables: $\exs \rightarrow \out$ and $\out \rightarrow \exs$. In the first model (Figure~\ref{subfig:expected_direction}), which we term $\model_\dir$, the causal relationship is in the expected direction, i.e., from exposure to outcome. In the alternate model (Figure~\ref{subfig:reverse_direction}), which we term $\model_\rev$, the causal relationship is in the opposite direction, which means we are dealing with reverse causation for the exposure-outcome association. Mendelian randomization rules out any other orderings for $(\iv_j, \exs)$ and $(\iv_j, \out)$ $\forall j \in \{1, 2, ... , J\}$, as the genotype assignment must have temporally preceded the other two variables. 

	The marginal likelihood for these models is also difficult to compute, with the Gaussian mixture prior on the parameters leading to a multimodal posterior. To solve our problem, we have used the \textsc{PolyChord}~\cite{handley_polychord_2015, handley_polychord_2015-1} algorithm, which employs a nested sampling scheme, for computing the (log-)evidences of the two models. We then estimate the causal effect in both directions and combine the results by weighting them appropriately with the model evidence. This combined result properly reflects the uncertainty of not knowing the direction of the causal relationship.
	
	If we have background knowledge that can be used to express a prior preference for one of the two models, we can easily incorporate it into our framework. We simply have to change the prior probabilities for each model. For example, if we know that $\exs$ precedes $\out$ temporally, we can eliminate the possibility of reverse causation. In that case, $p(\model_\dir) = 1$, and hence $p(\model_\dir \given \data) = 1$, so we are only considering the `expected direction' model. This is equivalent to the implicit assumption that the causal link is from exposure to outcome, which is often made in MR studies.

	\subsection{Posterior inference} 	
	
	\begin{wrapfigure}{r}{0.6\linewidth}
		\centering
		\includegraphics[width=\linewidth]{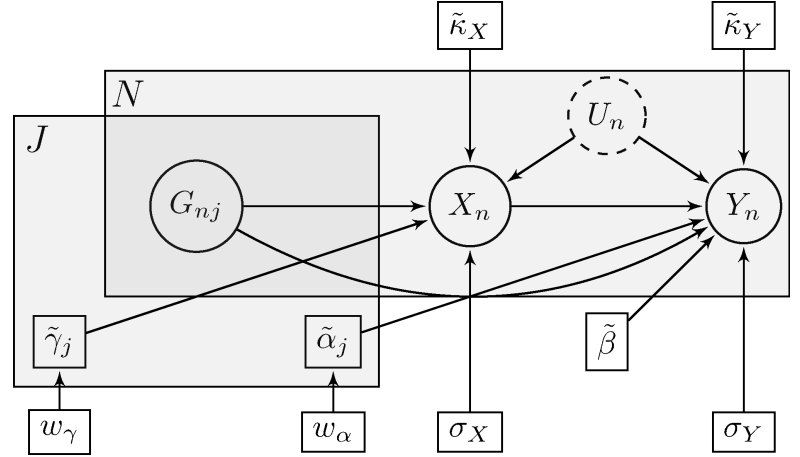}
		\caption{Compact description of the Bayesian inference process using plate notation. The small rectangles indicate our model parameters, while the circles denote random variables (observed or unobserved). The gray superposed areas signify replication across the $J$ genetic variants and the $N$ data points. This is also suggested by the subscripts used for parameters and variables.} \label{fig:inference}
	\end{wrapfigure}
	
	We take as input the complete data set of $\nobs$ i.i.d. observations $\data$ or the sample covariance matrix $\bfS$, which is a sufficient statistic for our model (see Subsection~\ref{ssec:summary_data}). As discussed in the previous section,  we fit the model described in Figure~\ref{fig:model} twice by considering both directions for the causal link between $\exs$ and $\out$, namely $\exs \to \out$ (Figure~\ref{subfig:expected_direction}) and $\out \to \exs$ (Figure~\ref{subfig:reverse_direction}). We first assume a causal link from exposure to outcome and perform Bayesian inference to derive the posterior distribution of $\cedir$. We then consider the reverse direction of causation and perform Bayesian inference to obtain the posterior distribution of $\cerev$.
	
	The Bayesian inference process is described in Figure~\ref{fig:inference} for the expected direction of the causal link. For the reverse direction, we simply switch the positions of $\exs$ and $\out$. Given our choice of prior distributions for the rescaled parameters (see Subsection~\ref{ssec:prior}), we draw $K$ samples from the full posterior distribution: $$ p(\sparm \given \data) \propto \lik_\data(\sparm) p(\sparm),$$ where $\sparm = (\rsc{\plyv}, \rsc{\ce}, \rsc{\isv}, \sigma_\exs, \sigma_\out, \rsc{\cc}_\exs, \rsc{\cc}_\out)$. After sampling, we revert our parameter of interest to the original scale: $$\ce^{(k)} = \frac{\sigma_\out^{(k)}}{\sigma_\exs^{(k)}} \rsc{\ce}^{(k)}, \forall k \in \natseq{K}.$$ 

	Finally, we compute the evidence for both directions, i.e., $p(\data \given \model_{\exs \to \out})$ and $p(\data \given \model_{\out \to \exs})$, and combine the posterior distributions over the parameters:
	\begin{equation}
	\begin{aligned}
	p(\cedir \given \data) &= p(\cedir \given \data, \model_\dir) \cdot p(\model_\dir) + \delta_0(\ce_\dir) \cdot p(\model_\rev) \\
	p(\cerev \given \data) &= p(\cerev \given \data, \model_\rev) \cdot p(\model_\rev) + \delta_0(\ce_\rev) \cdot p(\model_\dir) \\
	\end{aligned},
	\end{equation} where $\delta_0(\ce)$ is the Dirac delta function centered at zero.

	\subsection{Example: instrumental variable setting}
	
\begin{figure}[H]
		\begin{minipage}{0.49\linewidth}
			\centering
			\includegraphics[width=0.9\linewidth]{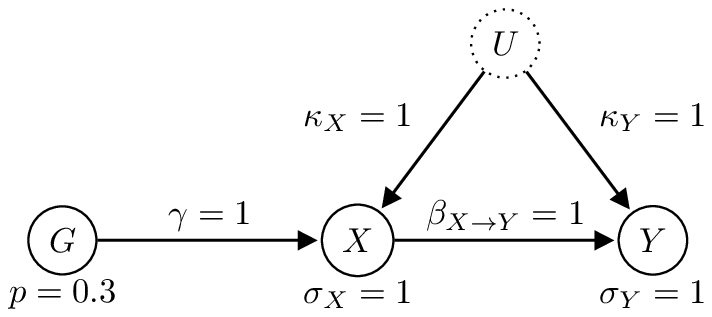}
			\caption{Generating model where the causal relationship is from $\exs$ to $\out$ (the expected direction). This model satisfies all three instrumental variable assumptions.} \label{fig:expected_direction_example}
		\end{minipage} \hfill
		\begin{minipage}{0.49\linewidth}
			\centering
			\includegraphics[width=0.9\linewidth]{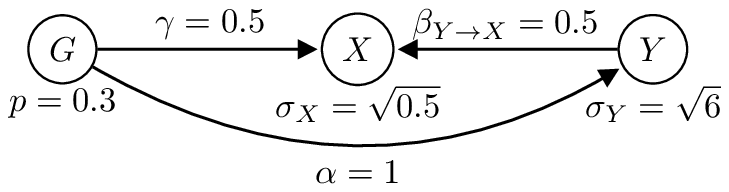}
			\caption{Alternative generating model, statistically equivalent to the one in Figure~\ref{fig:expected_direction_example}, where the causal relationship is from $\out$ to $\exs$ (the reverse direction).} \label{fig:alternative_IV}
		\end{minipage}
	\end{figure}

	We first consider a simple generating instrumental variable model with one genetic instrument (see Figure~\ref{fig:expected_direction_example}) to illustrate the model comparison and averaging procedure. In this example, the instrument strength $\is$, the causal effect $\ce$ and the confounding effect $\cc_\exs \cc_\out$ are of the same size ($\is = \ce =  \cc_\exs \cc_\out = 1$). We have purposefully chosen unrealistic structural parameter values for ease of illustration. Note, however, that the parameters can be scaled appropriately to reflect more realistic values. The genetic variant $\iv$ has effect allele frequency $\eaf = 0.3$. Assuming $\iv$ follows a binomial distribution with two allele copies, then $\Var{\iv} = 2 \cdot \eaf \cdot (1 - \eaf) = 0.42$. We also set $\mu_\exs = \mu_\out = 0$ and $\sigma_\exs = \sigma_\out = 1$. We generated 10000 samples from this model, resulting in the following sample covariance and correlation over the observed variables:%
	$$ \bSigma = \begin{bmatrix}
	0.421 & 0.434 & 0.441 \\
	0.434 & 2.447 & 3.439 \\
	0.441 & 3.439 & 6.404 \\
	\end{bmatrix}; \qquad \bfC = \begin{bmatrix} 
	1.000 & 0.427 & 0.268 \\
	0.427 & 1.000 & 0.869 \\
	0.268 & 0.869 & 1.000 \\
	\end{bmatrix}.$$
	
	We now fit models $\model_\dir$ (Figure~\ref{subfig:expected_direction}) and $\model_\rev$ (Figure~\ref{subfig:reverse_direction}) to the generated data and compute their log-evidence using a mixture Gaussian prior on each of the rescaled structural parameters ($\tilde{\is}, \tilde{\ply}, \tilde{\ce}, \tilde{\cc}_\exs, \tilde{\cc}_\out$), as described in Section~\ref{sec:model}. We fixed the hyperparameters of the Gaussian mixture prior to $\wt = 0.5, \tau = 1, \lambda = 0.01$ for each structural parameter. We obtained a Bayes factor of $\dfrac{p(\data \given \model_\dir)}{p(\data \given \model_\rev)} = 1.55$ in favor of the expected direction, which means that the data is 1.55 (95\% CI [0.959, 2.505]) times more likely to have been generated from $\model_\dir$ than $\model_\rev$. If we have no a-priori preference for $\model_\dir$ over $\model_\rev$ or vice versa, then the \textit{prior odds} are $\dfrac{p(\model_\dir)}{p(\model_\rev)} = 1$ and the corresponding \textit{posterior odds} are $\dfrac{p(\model_\dir \given \data)}{p(\model_\rev \given \data)} = 1.55 \; .$
	
	This result is perhaps surprising at first glance, as it implies that there is at least a one in three chance that the causal relationship is in the reverse direction. However, when we allow for pleiotropy, it becomes possible to fit a similarly sparse model where the $\exs \to \out$ edge is reversed. Such a model is illustrated in Figure~\ref{fig:alternative_IV}. Note that data generated from either the model in Figure~\ref{fig:expected_direction_example} or the one in Figure~\ref{fig:alternative_IV} have the same underlying probability distribution over the observed variables.	When we incorporate the assumption that pleiotropic effects are `weak' (close to zero), then the preference for $\model_\dir$ over $\model_\rev$ becomes very strong. We can express this preference in the prior for $\ply$, by setting $\wt = 0$ in Equation~\eqref{eqn:gauss_mix_prior}, which is equivalent to putting a strongly informative (low variance) Gaussian prior on the size of the pleiotropic effect. Note that this a weaker assumption than \texttt{IV3} ($\ply = 0$), which is equivalent to setting $\wt = \lambda = 0$ in the Gaussian mixture prior from Equation~\eqref{eqn:gauss_mix_prior}. After appropriately changing the prior of $\ply$ to reflect the assumption of `weak' pleiotropy by setting the hyperparameter $\wt$ to 0.0 instead of 0.5, the Bayes factor estimate becomes 100.36 (95\% CI [59.42, 169.51]), which translates into a probability that the model was $\model_\rev$ given the data of less than 1\%.

	\subsection{Example: near-conditional independence (near-LCD)} \label{ssec:nearLCD}
	
\begin{figure}[H]
		\begin{minipage}{0.49\linewidth}
			\centering
			\includegraphics[width=0.9\linewidth]{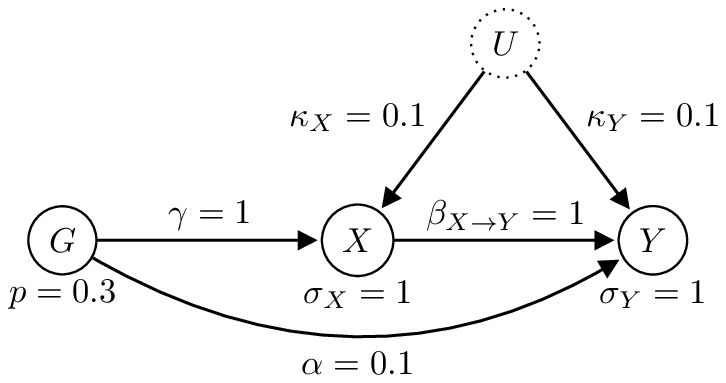}
			\caption{Generating model where the causal relationship is from $\exs$ to $\out$ ($\ce_\dir = 1$) and in which there is weak confounding ($\cc_\exs = \cc_\out = 0.1$) and pleiotropy ($\ply = 0.1$). There is a weak dependence $\iv \nindep \out \given \exs$ that results in a mild violation of the \texttt{IV3} assumption.} \label{fig:expected_direction_example_nearLCD}
		\end{minipage} \hfill
		\begin{minipage}{0.49\linewidth}
			\includegraphics[width=\linewidth]{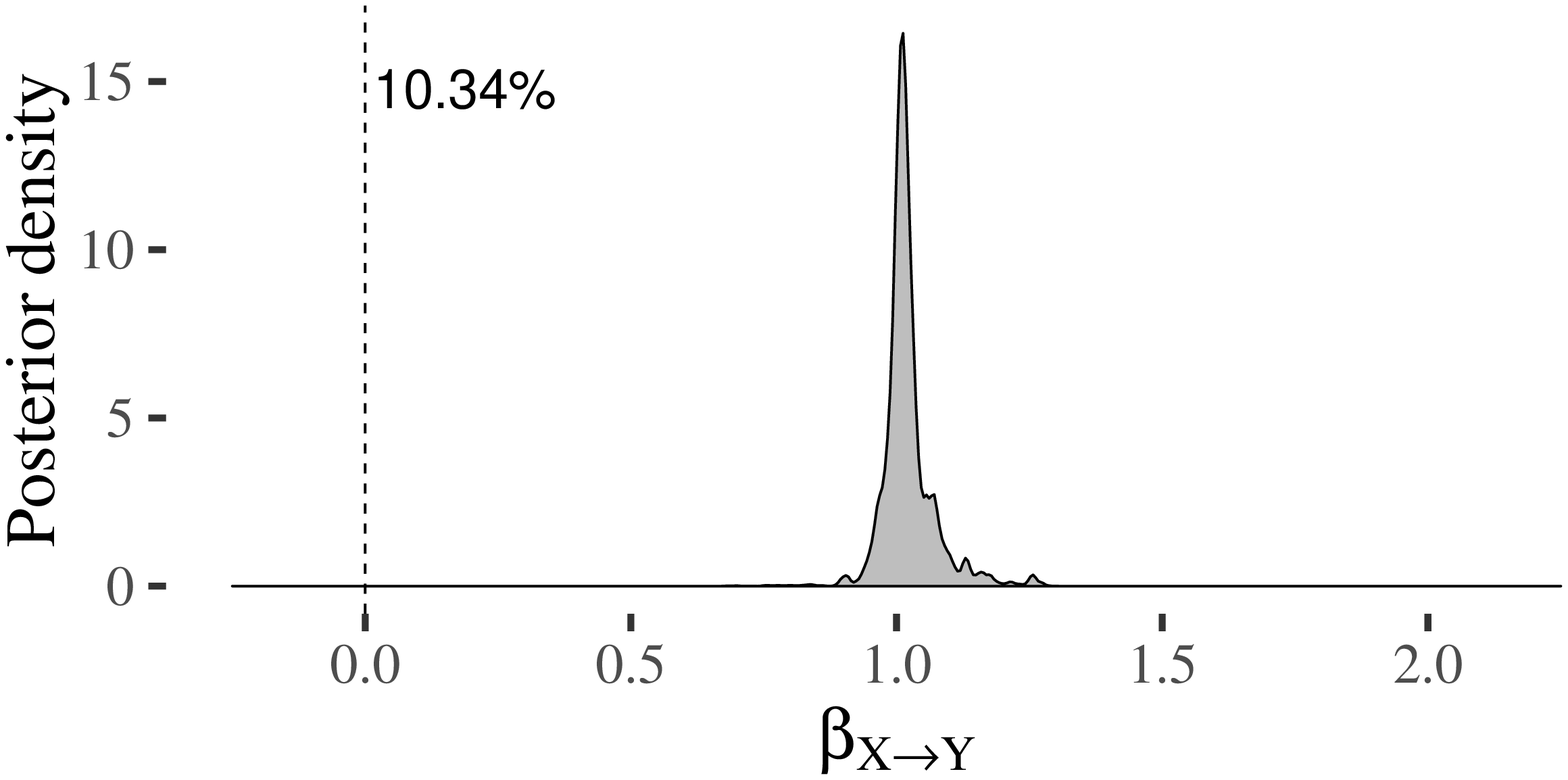}
			\caption{Estimated posterior density of the causal effect $\ce_{\exs \rightarrow \out}$ for the data generated from the model in Figure~\ref{fig:expected_direction_example_nearLCD}. The dashed vertical line at zero indicates the estimated probability of the reverse direction (Figure~\ref{subfig:reverse_direction}), in which case $\ce_\dir = 0$ since there is no causal link from $\exs$ to $\out$.} \label{fig:nearLCD_causal_effect}
		\end{minipage}
	\end{figure}

	We now look at an example where the confounding and pleiotropic effects are weak relative to the instrument strength and to the exposure-outcome causal effect. The generating model is depicted in Figure~\ref{fig:expected_direction_example_nearLCD}. We call this example near-LCD, because if we made the confounding and pleiotropic effects infinitely weak, i.e., $\ply = \cc = 0$, we would arrive at the LCD pattern from Figure~\ref{fig:lcd}. Again, we generated 10000 samples from the model, resulting in the following sample covariance and correlation matrices over the observed variables: %
	$$ \bSigma = \begin{bmatrix}
	0.425 & 0.430 & 0.481 \\
	0.430 & 1.477 & 1.538 \\
	0.481 & 1.538 & 2.645 \\
	\end{bmatrix} \bfC = \begin{bmatrix} 
	1.000 & 0.543 & 0.454 \\
	0.543 & 1.000 & 0.778 \\
	0.454 & 0.778 & 1.000 \\
	\end{bmatrix}.$$
	
	For the prior hyperparameters $\wt = 0.5, \tau = 1, \lambda = 0.01$, we obtained a Bayes Factor in favor of the expected direction of: $\dfrac{p(\model_\dir \given \data)}{p(\model_\rev \given \data)} = 5.90 \; .$ The confidence interval for the Bayes factor value (95\% CI $[3.68, 9.47]$) suggests that the preference for the expected direction is significant. This example shows that under mild violations of the $\texttt{IV3}$ assumption and given weak confounding effects, we can reliably indicate the correct causal direction using our method. Moreover, \textsc{BayesMR} appropriately computes the degree of uncertainty in the model selection, which increases for stronger confounding and/or pleiotropic effects. Finally, our approach results in reasonable posterior samples, which we show in Figure~\ref{fig:nearLCD_causal_effect}. We can see that most of the mass of the posterior is concentrated around the true value $\ce_\dir = 1$. The density has two components, given by the two possible causal directions. The dashed vertical line at zero indicates the estimated probability of the reverse direction, for which $\beta_{\exs \to \out} = 0$ since there is no causal link from $\exs$ to $\out$.

	\section{Simulation studies} \label{sec:simulation}
	
	\subsection{Estimation robustness}
	
	We imagined a typical MR scenario where there are a number ($J = 25$)  of genetic variants that may be used as instruments, but it is not known how many of them satisfy the IV assumptions. We analyzed the robustness of the causal effect estimate produced by our method to the presence of pleiotropy. We first treated the case where all variants were valid instruments. We then added pleiotropic effects to the genetic variants, such that only 80\%, 60\%, 40\%, 20\%, and finally 0\% of them were valid instruments. We generated data according to the Bayesian model specification in Section~\ref{sec:model} (see Figure~\ref{fig:model}). The effect allele frequencies of each genetic variant were randomly simulated from a uniform distribution. We simulated the instrument strengths ($\isv$) from a normal distribution with standard deviation $\sigma_\is = 0.1$. The average absolute strength of the instruments is the mean of the corresponding half-normal distribution: $\EV{|\is|} = \frac{2 \sigma_\is}{\sqrt{2 \pi}} \approx 0.08$. We simulated the intrinsic noise of the exposure and outcome from a half-normal distribution with standard deviation equal to one, which means that $\EV{\sigma_\exs} = \EV{\sigma_\out} = \frac{2}{\sqrt{2 \pi}} \approx 0.8$. The causal effect from $\exs$ to $\out$ was set at $\ce = 1$. The pleiotropic effects introduced were simulated from a normal distribution with mean zero and standard deviation $\sigma_\ply = 0.1$, which means that they have a similar magnitude to that of the instrument strengths. This type of pleiotropy is referred to as \textit{balanced} in the literature.~\cite{bowden_framework_2017} %
	
	\begin{figure}[!htb]
		\begin{minipage}{0.55\linewidth}
			\includegraphics[width=\linewidth]{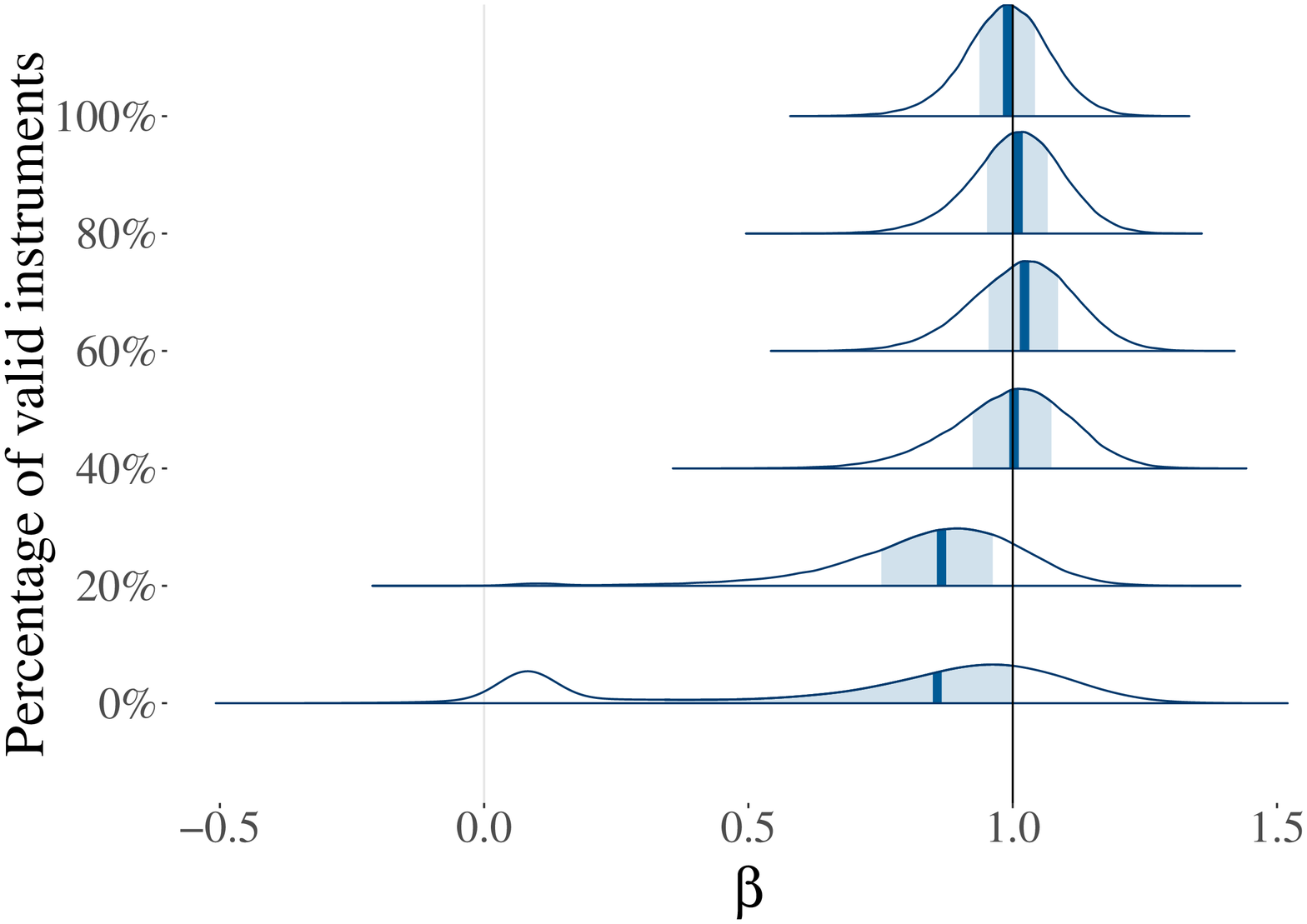}
			\caption{The effect of introducing pleiotropic effects on the posterior estimates is that the distribution moves away from the true value $\ce = 1$. The shaded area in each posterior distribution corresponds to the 50\% posterior uncertainty (credible) interval, with the posterior median in the center depicted with a vertical line. In the worst case scenario, where no genetic variant is a valid instrument, we observe the appearance of a second mode of the distribution, which is close to zero. This mode corresponds to the model explanation of the data where there is a `weak' causal effect from exposure to outcome. We notice, however, that the posterior distribution progression is gradual, thereby showcasing the robustness of \textsc{BayesMR} to the presence of pleiotropy. When only 40\% of the genetic variants were valid instruments, the posterior distribution remained robustly centered around $\ce = 1$. Even when none of the genetic variants satisfied the IV assumptions, a significant proportion of the probability mass could be found around the true value. } \label{fig:posterior_ply}
		\end{minipage} \hfill
		\begin{minipage}{0.4\linewidth}
			\includegraphics[width=\linewidth]{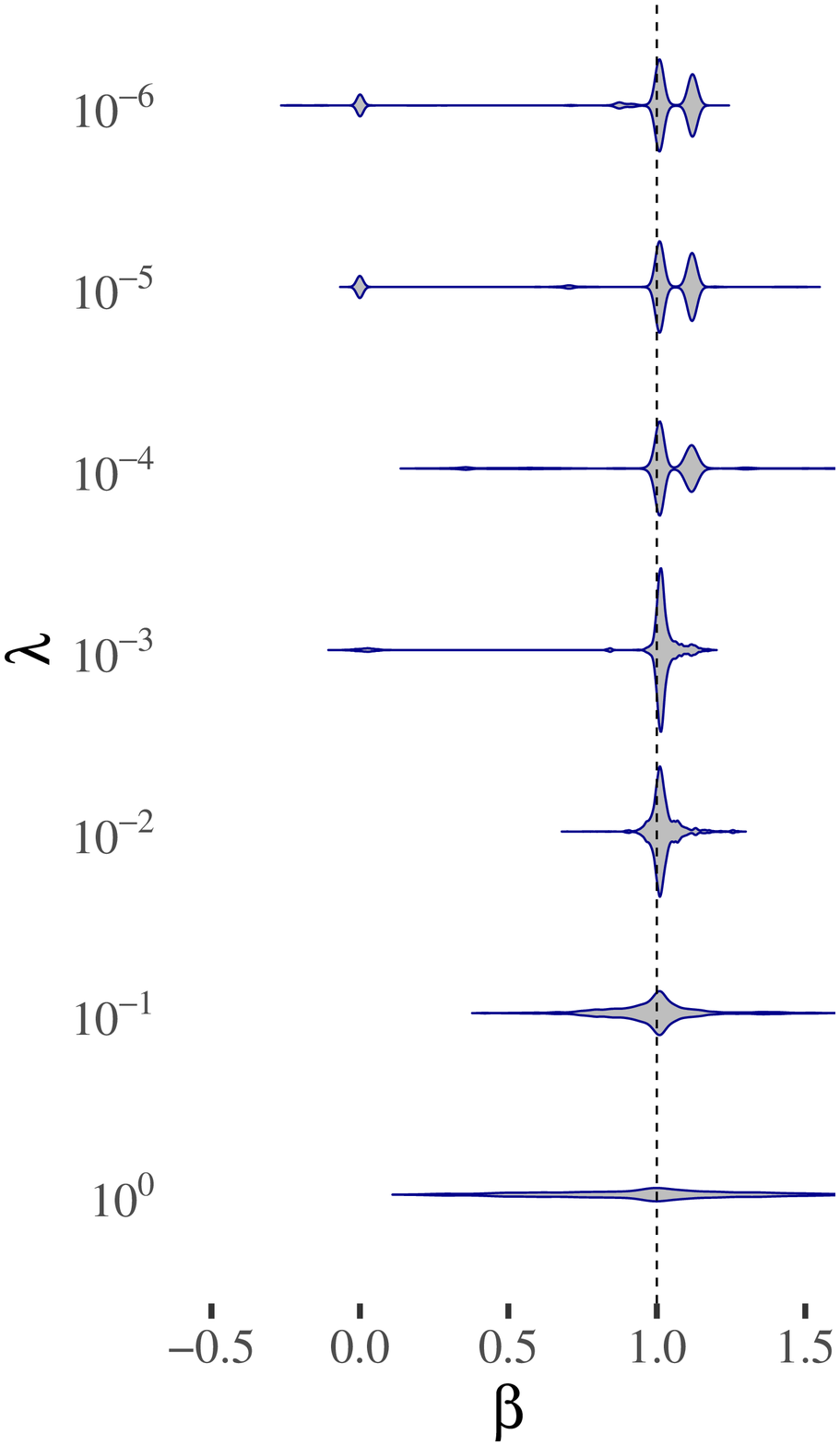}
			\caption{Posterior distribution of $\ce$ for different hyperparameter settings (only $\lambda$ is varied) in the near-LCD scenario (Figure~\ref{fig:expected_direction_example_nearLCD}). The true value for $\beta$ is depicted by a dashed vertical line.} \label{fig:spike_robustness}
		\end{minipage}
	\end{figure}
	For the inference, we set the prior hyperparameters to $\tau = 1, \lambda = 0.01$. The weights indicating the proportion of relevant genetic effects are learned from the data ($\wt_{\isv}$ and $\wt_{\plyv}$), while for the other parameters having a Gaussian mixture prior, the weight hyperparameter is set to $\wt = 0.5$. In Figure~\ref{fig:posterior_ply} we show how the causal effect estimate is affected as fewer and fewer of the genetic variants are valid instruments. We notice that even when only 40\% of the genetic variants satisfied the IV assumptions, the posterior distribution remained robustly centered around the true value $\ce = 1$. When none of the genetic variants were valid instrumental variables, this was properly reflected in the uncertainty of the posterior distribution. In that particular case, the posterior became bimodal, suggesting two explanations for the observed associations, one where there is a strong causal link between exposure and outcome (the peak around one) and one when the associations are due to a combination of pleiotropic genetic effects and confounding (the peak close to zero).

	\subsection{Sensitivity to prior hyperparameters}
	
	In this simulation, we analyzed how the posterior computed by \textsc{BayesMR} is influenced by the choice of hyperparameters in the Gaussian mixture prior. For this purpose, we again considered the near-LCD scenario introduced in Subsection~\ref{ssec:nearLCD} (see Figure~\ref{fig:expected_direction_example_nearLCD}). We fixed $\tau = 1$ and varied the parameter $\lambda$ in equally sized steps on the logarithmic scale. In this configuration, the higher variance (`slab') component of the Gaussian mixture prior has fixed width, while the lower variance (`spike') component's width depends on $\lambda$. The mixture weights were fixed at 0.5. For each of the structural parameters on which we put a Gaussian mixture parameter, namely $\gamma, \alpha, \beta, \kappa_\exs, \kappa_\out$, we used the same hyperparameter configuration. The posterior distributions obtained for each value of $\lambda$ are shown in Figure~\ref{fig:spike_robustness}. We notice that for every different spike width, the posterior centers around the correct value $\beta = 1$. For narrower spikes, the posterior becomes multimodal, indicating other possible sparse solutions. For example, the mode centered around $\beta = 1.1$ corresponds to the `weak pleiotropy' solution, where $\ply \approx 0$. The mode centered at zero, on the other hand, indicates that there is a very small chance that there is no causal effect from $\exs$ to $\out$.
	
	The obtained posterior is sensitive to the hyperparameter choice, as reflected by the different mode heights in the multimodal posterior. This is due to the fact that the prior incorporates our belief regarding the magnitude of `weak' and `strong' effects. For instance, the appearance of the second mode around $1.1$ for smaller values of $\lambda$ can be explained by the change in the prior belief that the pleiotropic effect, which is of magnitude $\alpha = 0.1$, is relevant or irrelevant. For $\lambda = 10^{-2}$, an effect of magnitude $0.1$ has a prior odds of coming from the `spike' (irrelevant interaction) instead of the `slab' (relevant interaction) roughly equal to $6.1$. This means that for $\lambda = 10^{-2}$, it is easy to `fit' a parameter of that magnitude into the `spike' component, where it is assigned a relatively high prior probability which is similar to the prior probability other small values, including zero. In that case, a solution close to the ground truth, where the confounding and pleiotropy parameters are fit to irrelevant values, will have a high posterior probability. This is reflected by the posterior mode for $\beta$ around $1$ in Figure~\ref{fig:spike_robustness}. In contrast, when $\lambda = 10^{-4}$, then the prior odds for an effect of magnitude $0.1$ to come from the `spike', instead of the `slab', is of order $10^{-20}$. This means that an effect of magnitude $0.1$ is a-priori extremely likely to be considered relevant. Because of the sparsifying Gaussian mixture prior, it is however much more advantageous to fit an irrelevant horizontal pleiotropy parameter instead of a relevant one, even though the ground truth is $\alpha = 0.1$. This means that the statistically equivalent (resulting in the same covariance matrix) solution where $\alpha = 0$ and $\beta = 1.1$ now has a high enough posterior probability, which leads to the appearance of the mode around $\beta = 1.1$.
	
	Even though, as we have shown, the posterior distribution is driven by the choice of prior hyperparameters, our approach also enables the researcher to easily incorporate useful background knowledge, such as previously established weak or strong interactions, into the system. The advantage lies in the intuitive interpretation of the hyperparameters for the Gaussian mixture prior (see Subsection~\ref{ssec:prior}).

	\subsection{Model averaging robustness}
	
	In this experiment, we considered a bidirectional Mendelian randomization setting (see Subsection~\ref{ssec:reverse_causation}), where we cannot exclude the presence of pleiotropic effects. For simplicity, we considered a single genetic variant $\iv_\exs$ known to be associated with $\exs$ and a single genetic variant $\iv_\out$ associated with $\out$. The true causal direction in the generating model was from $\exs$ to $\out$. This generating model is described by the set of equations:
	\begin{equation} \label{eqn:bidirectional_MR}
	\begin{aligned}
	\exs &:= \is_\exs \iv_\exs + \ply_\exs \iv_\out + \cc_\exs \conf + \erre_\exs \\		
	\out &:= \ply_\out \iv_\exs + \is_\out \iv_\out + \ce \exs + \cc_\out \conf + \erre_\out, \\
	\end{aligned}
	\end{equation} where $\is_\exs$ and $\is_\out$ are the instrument strengths (for $\exs$ and for $\out$ respectively), $\ply_\exs$ and $\ply_\out$ are the pleiotropic (secondary) effects, $\cc_\exs$ and $\cc_\out$ are the confounding coefficients and $\ce$ is the causal effect of interest.
	
	In our simulation, the genetic variants were strongly associated with their respective phenotypes: $\is_\exs = \is_\out = 1$. The causal effect parameter was set to $\ce_{\exs \to \out} = 1$. We varied the pleiotropic effects for each simulation by setting $\ply_\exs = \ply_\out = \delta$, with $\delta \in \{-0.5, -0.4, -0.3, ..., 0.5\}$. At the same time, we considered strong confounding: $\cc_\exs = \cc_\out = 1$, while the error terms were normally distributed with dispersion given by $\sigma_\exs = \sigma_\out = 1$. We generated 10000 samples from the above structural equation model with these parameter configurations. 
	
	We used \textsc{BayesMR} to determine how much more likely the $\exs \rightarrow \out$ link is than $\out \rightarrow \exs$ and to produce a robust causal estimate by combining the estimates for both causal directions. The results of a bidirectional MR analysis, which amounts to computing the Wald ratio from Equation~\eqref{eqn:wald_ratio} for both causal directions, are shown in Table~\ref{tab:bidirectional_MR} for each value of $\delta$. When $\delta = 0$, the conditions are ideal for the application of bidirectional Mendelian randomization, since both genetic variants constitute valid instruments for their respective phenotypes. In this case, bidirectional MR would yield an estimate of $\hat{\ce}^{IV} = 1.006$ (95\% CI [0.934, 1.078]) for the $\exs \to \out$ causal direction and $\hat{\ce}^{IV} = 0.018$ (95\% CI [-0.025, 0.062]) for the $\out \to \exs$ causal direction. Since the causal effect estimate is very close to zero for the link $\out \to \exs$, the inferred direction would be $\exs \to \out$, for which the causal effect estimate is close to the true value.
	\begin{table}[H]
		\begin{center}
			\begin{tabular}{r|rrrr|c}
				\multicolumn{1}{c}{$\delta$}&\multicolumn{1}{c}{$\hat{\ce}^{IV}_{X \to Y}$}&\multicolumn{1}{c}{$\hat{\sigma}^{IV}_{X \to Y}$}&\multicolumn{1}{c}{$\hat{\ce}^{IV}_{Y \to X}$}&\multicolumn{1}{c}{$\hat{\sigma}^{IV}_{Y \to X}$}&\multicolumn{1}{c}{$\hat{p}(\model_{X \to Y} \given \data)$}\tabularnewline
				\hline
				$-0.5$&$0.517$&$0.0346$&$-0.8715$&$0.0415$&$0.167$\tabularnewline
				$-0.4$&$0.615$&$0.0348$&$-0.5853$&$0.0352$&$0.211$\tabularnewline
				$-0.3$&$0.712$&$0.0350$&$-0.3746$&$0.0305$&$0.516$\tabularnewline
				$-0.2$&$0.810$&$0.0353$&$-0.2130$&$0.0270$&$0.632$\tabularnewline
				$-0.1$&$0.908$&$0.0356$&$-0.0851$&$0.0242$&$0.894$\tabularnewline
				$ 0.0$&$1.006$&$0.0359$&$ 0.0187$&$0.0219$&$0.907$\tabularnewline
				$ 0.1$&$1.105$&$0.0362$&$ 0.1045$&$0.0200$&$0.774$\tabularnewline
				$ 0.2$&$1.204$&$0.0366$&$ 0.1767$&$0.0184$&$0.864$\tabularnewline
				$ 0.3$&$1.303$&$0.0370$&$ 0.2383$&$0.0170$&$0.803$\tabularnewline
				$ 0.4$&$1.402$&$0.0374$&$ 0.2914$&$0.0159$&$0.750$\tabularnewline
				$ 0.5$&$1.501$&$0.0378$&$ 0.3377$&$0.0149$&$0.774$\tabularnewline
				\hline
		\end{tabular}\end{center}
		\caption{We show the results of performing a simple bidirectional Mendelian randomization analysis for various degrees of pleiotropy. In the first step, we used $\iv_\exs$ as an instrument for $\exs$ to estimate the causal effect $\beta_{\exs \to \out}$ (the correct direction). We then used $\iv_\out$ as an instrument for $\out$ to estimate the causal effect $\beta_{\out \to \exs}$ (the wrong direction). We compared these estimates against our posterior probability estimate of $\model_{X \to Y}$ given the data, in which analysis we used both instruments concomitantly.} \label{tab:bidirectional_MR}
	\end{table}
		
	In the presence of pleiotropic effects ($\delta \neq 0$), the bidirectional MR analysis is no longer suitable, since it is possible to arrive at a nonzero causal effect in both directions. In Table~\ref{tab:bidirectional_MR}, we obtain a nonzero causal effect for both $\exs \to \out$ and $\out \to \exs$ anytime $\delta \neq 0$. Furthermore, it is possible that the causal effect estimated in the wrong direction ($\out \to \exs$) is stronger in magnitude than the causal effect estimated in the correct direction ($\exs \to \out$). This shows that bidirectional MR is not a reliable means of inferring the causal direction when dealing with pleiotropic genetic variants. Our approach, on the other hand, was robust to the presence of weak pleiotropic effects. In the last column of Table~\ref{tab:bidirectional_MR}, we observe a significant preference for the correct direction when the pleiotropic effects are small.

	\subsection{Sensitivity to nonlinearity and nonnormality}	
	
	Our assumed Bayesian model, as described in Section~\ref{sec:model}, entails a linear relationship between the exposure $\exs$ and the outcome $\out$, as well as a conditional Gaussian distribution given the instrument values, $p(\exs, \out \given \ivv)$. In this section, we evaluate the robustness of \textsc{BayesMR} in situations where the parametric assumptions of linearity and normality are violated. We revisit the near-LCD scenario from Subsection~\ref{ssec:nearLCD}, where our approach is able to recover the correct direction of the causal relationship with high probability and to produce a reasonable estimate for the causal effect, and use it as the starting point for our sensitivity analysis.
	
	\begin{minipage}[t]{0.48\linewidth}
		\begin{table}[H]
			\footnotesize
				\begin{tabular}{c|cccc|c}
					\multicolumn{1}{c}{} & \multicolumn{4}{c}{$\hat{\ce}_{\exs \to \out}$} & \multicolumn{1}{c}{} \\
					\multicolumn{1}{c}{A}&\multicolumn{1}{c}{Q1}&\multicolumn{1}{c}{Median}&\multicolumn{1}{c}{Mean}&\multicolumn{1}{c}{Q3}&\multicolumn{1}{c}{$\hat{p}(\model_{X \to Y} \given \data)$}\tabularnewline
					\hline
					$0.5$&$0.260$&$0.289$&$0.276$&$0.309$&$[0.744, 0.880]$\tabularnewline
					$1$&$0.494$&$0.511$&$0.500$&$0.530$&$[0.817, 0.919]$\tabularnewline
					$2$&$0.739$&$0.758$&$0.737$&$0.786$&$[0.821, 0.921]$\tabularnewline
					$4$&$0.906$&$0.921$&$0.927$&$0.940$&$[0.780, 0.900]$\tabularnewline
					$8$&$0.970$&$0.985$&$0.992$&$1.003$&$[0.815, 0.918]$\tabularnewline
					$16$&$0.990$&$1.006$&$1.011$&$1.025$&$[0.776, 0.899]$\tabularnewline
					$32$&$0.991$&$1.009$&$0.996$&$1.028$&$[0.827, 0.924]$\tabularnewline
					\hline
			\end{tabular}
			\caption{We show the results of running \textsc{BayesMR} on data generated from the model in Figure~\ref{fig:nearLCD_causal_effect} for various degrees of nonlinearity, as parametrized by $A$ (lower $A$ corresponds to stronger nonlinearity). We report the 95\% CI for the estimated probability that the direction of the causal link is $\exs \to \out$, as opposed to $\out \to \exs$. For the causal effect estimate $\hat{\ce}_{\exs \to \out}$, we report summary measures, including the first (Q1) and third quartile (Q3).} \label{tab:sensitivity_nonlinear}
		\end{table}
	\end{minipage} \hfill %
	\begin{minipage}[t]{0.48\linewidth}
		\begin{table}[H]
			\footnotesize
			\begin{tabular}{c|cccc|c}
					\multicolumn{1}{c}{} & \multicolumn{4}{c}{$\hat{\ce}_{\exs \to \out}$} & \multicolumn{1}{c}{} \\
					\multicolumn{1}{c}{$\nu$}&\multicolumn{1}{c}{Q1}&\multicolumn{1}{c}{Median}&\multicolumn{1}{c}{Mean}&\multicolumn{1}{c}{Q3}&\multicolumn{1}{c}{$\hat{p}(\model_{X \to Y} \given \data)$}\tabularnewline
					\hline
					$1$&$-1.541$&$0.001$&$0.294$&$1.998$&$[0.635, 0.823]$\tabularnewline
					$2$&$~0.904$&$1.010$&$0.915$&$1.066$&$[0.743, 0.876]$\tabularnewline
					$4$&$~0.990$&$1.019$&$0.990$&$1.044$&$[0.879, 0.948]$\tabularnewline
					$8$&$~1.013$&$1.032$&$1.021$&$1.053$&$[0.809, 0.914]$\tabularnewline
					$16$&$~0.998$&$1.015$&$1.017$&$1.033$&$[0.828, 0.924]$\tabularnewline
					$32$&$~0.983$&$0.999$&$0.980$&$1.017$&$[0.852, 0.936]$\tabularnewline
					$64$&$~1.007$&$1.024$&$1.025$&$1.043$&$[0.800, 0.913]$\tabularnewline
					\hline
			\end{tabular}
			\caption{We show the results of running \textsc{BayesMR} on data generated from the model in Figure~\ref{fig:nearLCD_causal_effect} for various degrees of nonnormality, as parametrized by the number of degrees of freedom of the $t$-distributed noise (lower $\nu$ corresponds to stronger nonnormality). We report the 95\% CI for the estimated probability that the direction of the causal link is $\exs \to \out$, as opposed to $\out \to \exs$. For the causal effect estimate $\hat{\ce}_{\exs \to \out}$, we report summary measures, including the first (Q1) and third quartile (Q3).} \label{tab:sensitivity_nongaussian}
		\end{table}
	\end{minipage}
	
	We first explore the situation in which the relationship between exposure and outcome is nonlinear. To make the results comparable to the linear case, we have simulated data where the linear term in the structural equation of $\out$ from ~\eqref{eqn:linear_model} is replaced by $A \tanh(\frac{\exs}{A})$, where the parameter $A$ controls the degree of nonlinearity. In the limit $A \to \infty$, $\tanh(\frac{\exs}{A}) \to \frac{\exs}{A}$, so we recover linearity. Furthermore, the coefficient of the linear tangent approximation at $\exs = 0$ has the same value (one) as the structural coefficient $\ce$ in the near-LCD scenario. We keep the other model parameters as in the example in Subsection~\ref{ssec:nearLCD}. The results in Table~\ref{tab:sensitivity_nonlinear} show that despite a nonlinear $\exs-\out$ dependence, we are still able to detect the causal effect and its sign. Moreover, the method returns a strong preference for the correct direction for all values of $A$. The causal effect estimate $\hat{\ce}_{\exs \to \out}$ remains robust against small to medium deviations from the linear case. It is important to note that when the linearity assumption does not hold, then the estimate returned by our method cannot be interpreted as the effect of an increase in the exposure for an individual, but represents an average causal effect across the population.~\cite{burgess_instrumental_2014} In case the exposure-outcome relationship is non-monotonic, for instance if it is U-shaped, then it is no longer guaranteed that the method can infer any causal effect. One possible solution then would be to perform a piecewise Mendelian randomization analysis, as suggested by Burgess et al.~\cite{burgess_instrumental_2014} The idea is to stratify on the ``IV-free'' exposure distribution so that the localized exposure-outcome relation is approximately linear and then run \textsc{BayesMR} on each stratum.

	In the second simulation, we analyze the effects of including non-Gaussian noise in the generating model. To make the results comparable to the Gaussian case, we have simulated data where the noise term $\err_\out$ in the structural equation of $\out$ from~\eqref{eqn:linear_model} is distributed according to Student's $t$-distribution. Here, the $t$-distribution's number of degrees of freedom $\nu$ controls the degree of normality. In the limit $\nu \to \infty$, the $t$-distribution becomes a standard normal distribution, so we recover normality. We keep the other model parameters as in the example in Subsection~\ref{ssec:nearLCD}. The results in Table~\ref{tab:sensitivity_nongaussian} show that the causal effect estimate for $\ce_{\exs \to \out}$ remains robust for a large range of values for $\nu$, corresponding to small to medium deviations from the linear case. Moreover, the method returns a strong preference for the correct direction of the causal link for all values of $\nu$. When $\nu = 1$, which corresponds to a $t$-distribution with undefined mean and variance, our approach can no longer detect the causal link from $\exs$ to $\out$, as the posterior distribution of the causal effect estimate becomes too broad. One potential solution for dealing with strong violations of the normality assumption would then be to adopt a non-Gaussian structural equation model, in the vein of Shimizu and Bollen.~\cite{shimizu_bayesian_2014}

	\section{Real-world applications} \label{sec:applications}
	
	In this section, we will showcase the potential of \textsc{BayesMR} by applying it to a number of real-world problems. We start, however, by explaining how our method can be applied when only summary data is available. In the first application, we focus on estimating the causal effect of birth weight on adult fasting glucose levels. In the second application, we analyze the effect of body mass index on the risk of Parkinson's disease. In the third and final application, we use our approach to examine the direction of the causal link between coffee consumption and cigarette smoking.

	\subsection{Using summary data} \label{ssec:summary_data}
	
	\textsc{BayesMR} requires as input the first and second-order moments of the observed data vector $\bfZ = (\ivv, \exs, \out)$. These moments are needed in the expression of the likelihood function and constitute the sufficient statistics with respect to our model (Section~\ref{sec:model}). We can easily compute estimates for these moments when individual-level data is available. However, we often only have access to summary statistics in the form of regression (beta) coefficients and their  standard errors. In this subsection,  we show how to derive approximations for the first and second-order moments from the summary data in order to obtain the required sufficient statistics for our method. We require (at least) the following summary statistics:
	\begin{itemize}
		\item $\hat{\eaf}_j$: the effect allele frequency (EAF) of $\iv_j$
		\item $\nac$: the number of allele copies (very often equal to two, since humans are diploid organisms)
		\item $\hat{\is}_j, \hat{\sigma}_{\hat{\is}_j}, \nobs_{\hat{\is}_j}$: measures of the gene-exposure association, including the coefficient obtained by regressing $\exs$ on $\iv_j$ , its standard error and the sample size 
		\item $\hat{\te}_j, \hat{\sigma}_{\hat{\te}_j}, \nobs_{\hat{\te}_j}$: measures of the gene-outcome association, including the coefficient obtained by regressing $\out$ on $\iv_j$, its standard error and the sample size
		\item $\hat{\ce}$: the coefficient obtained by regressing $\exs$ and $\out$ (observational exposure-outcome association)
	\end{itemize}
	
	Summary data on gene-exposure and gene-outcome associations from genome-wide association studies has become increasingly available, so we can typically get estimates for $\hat{\is}_j$, $\hat{\te}_j$ together with the associated standard errors and sample sizes. The effect allele frequency $\hat{\eaf}_j$ is usually also reported. In addition, we require a measure of the association between the exposure and the outcome ($\hat{\ce}$) to derive an estimate of $\Cov{\exs, \out}$. This estimate can be obtained from observational studies for determining potential risk factors for the outcome. 
	
	To derive the expected values and variances for the genetic variants, we employ the binomial distribution assumption by plugging in the EAF as the estimated success probability and using the appropriate formulas. We can assume without loss of generality that the mean parameters $\mu_\exs$ and $\mu_\out$ in Equation~\eqref{eqn:sem} are zero, as their location does not influence the regression slope. For estimating the second-order moments, we then employ the following well-known approximations from simple linear regression:
	\begin{equation*}
	\small
	\begin{aligned}
		\hat{\is}_j &\approx \frac{\Cov{\iv_j, \exs}}{\Var{\iv_j}} \\
		\hat{\te}_j &\approx \frac{\Cov{\iv_j, \out}}{\Var{\iv_j}} \\
		\hat{\ce} &\approx \frac{\Cov{\exs, \out}}{\Var{\exs}} \\
		\hat{\sigma}^2_{\hat{\is}_j} &\approx \frac{1}{N_\is} \left( \frac{\Var{\exs}}{\Var{\iv_j}} - \hat{\is}^2 \right) \\
		\hat{\sigma}^2_{\hat{\te}_j} &\approx \frac{1}{N_\te} \left( \frac{\Var{\out}}{\Var{\iv_j}} - \hat{\te}^2 \right)
	\end{aligned}
	\end{equation*}
	
	Note that these approximations also apply in a multivariate setting when the regressors are independent. We use these approximations to finally derive the following estimates for the moments from summary statistics:
	
	\begin{equation} 
	\small \label{eqn:summary_stat}
	\begin{aligned} 
		\EV{\iv_j} &\approx \nac \cdot \hat{\eaf}_j \; (= \widehat{\EV{\iv_j}}) \\
		\EV{X} &\approx \sum_j \widehat{\EV{\iv_j}} \cdot \hat{\is}_j \\
		\EV{Y} &\approx \sum_j \widehat{\EV{\iv_j}} \cdot \hat{\te}_j \\
	\end{aligned} \hskip 0.1\textwidth 
	\begin{aligned}
		\Var{\iv_j} &\approx \nac \cdot \hat{\eaf}_j \cdot (1 - \hat{\eaf}_j) \; (= \widehat{\Var{\iv_j}}) \\
		\Cov{\iv_j, \exs} &\approx \widehat{\Var{\iv_j}} \cdot \hat{\is}_j \\
		\Cov{\iv_j, \out} &\approx \widehat{\Var{\iv_j}} \cdot \hat{\te}_j \\
		\Var{\exs} &\approx \widehat{\Var{\iv_j}} \cdot (\hat{\is}_j^2 + \nobs_{\hat{\is}_j} \cdot \hat{\sigma}^2_{\hat{\is}_j}) \; (= \widehat{\Var{\exs}}) \\
		\Var{\out} &\approx \widehat{\Var{\iv_j}} \cdot (\hat{\te}_j^2 + \nobs_{\hat{\te}_j} \cdot \hat{\sigma}^2_{\hat{\te}_j}) \\
		\Cov{\exs, \out} &\approx \widehat{\Var{\exs}} \cdot \hat{\ce}
	\end{aligned}
	\end{equation}
	When we have information on multiple genetic variants, we obtain multiple estimates of $\Var{\exs}$ and $\Var{\out}$ in~\eqref{eqn:summary_stat}, in which case we use the average over the estimates. Our approach also requires specifying a sample size. Since the summary statistics are likely to be computed from different samples, we conservatively choose the minimum of their sizes as input to \textsc{BayesMR} in order not to overestimate the precision of the data. If the sample size for the exposure-outcome association measure is also available, we take it into consideration when calculating the minimum of the sample sizes.

	\subsection{Effect of birth weight on fasting glucose}
	
	Del Greco et al.~\cite{del_greco_detecting_2015} performed an IVW meta-analysis of MR Wald estimates (see Equations~\eqref{eqn:wald_ratio},~\eqref{eqn:ivw}) to analyze the relationship between low birth weight and adult fasting glucose levels. The authors chose seven genetic variants associated with birth weight to use as instruments in an MR analysis. The meta-analysis of the seven MR estimates suggested a significant protective effect of -0.155 $mmol/L$ (95\% CI [-0.233, -0.088]) reduction in adult fasting glucose level per standard deviation increase (484 g) in birth weight. In their analysis, Del Greco et al.~\cite{del_greco_detecting_2015} investigated the presence of pleiotropy by means of the between-instrument heterogeneity $Q$ test. They reported significant evidence of heterogeneity across instruments ($p = 0.03$), which they believe suggests that \texttt{IV3} might be violated for some of the genetic variants. The heterogeneity was primarily due to the variant \texttt{rs9883024} in gene \texttt{ADCY5}. However, even after removing said variant from the analysis, they obtained a significant negative effect estimate of -0.098 (95\% CI [-0.168, -0.027]).
	
	\begin{figure}[!htb]
		\begin{minipage}{0.49\linewidth}
			\includegraphics[width=\textwidth]{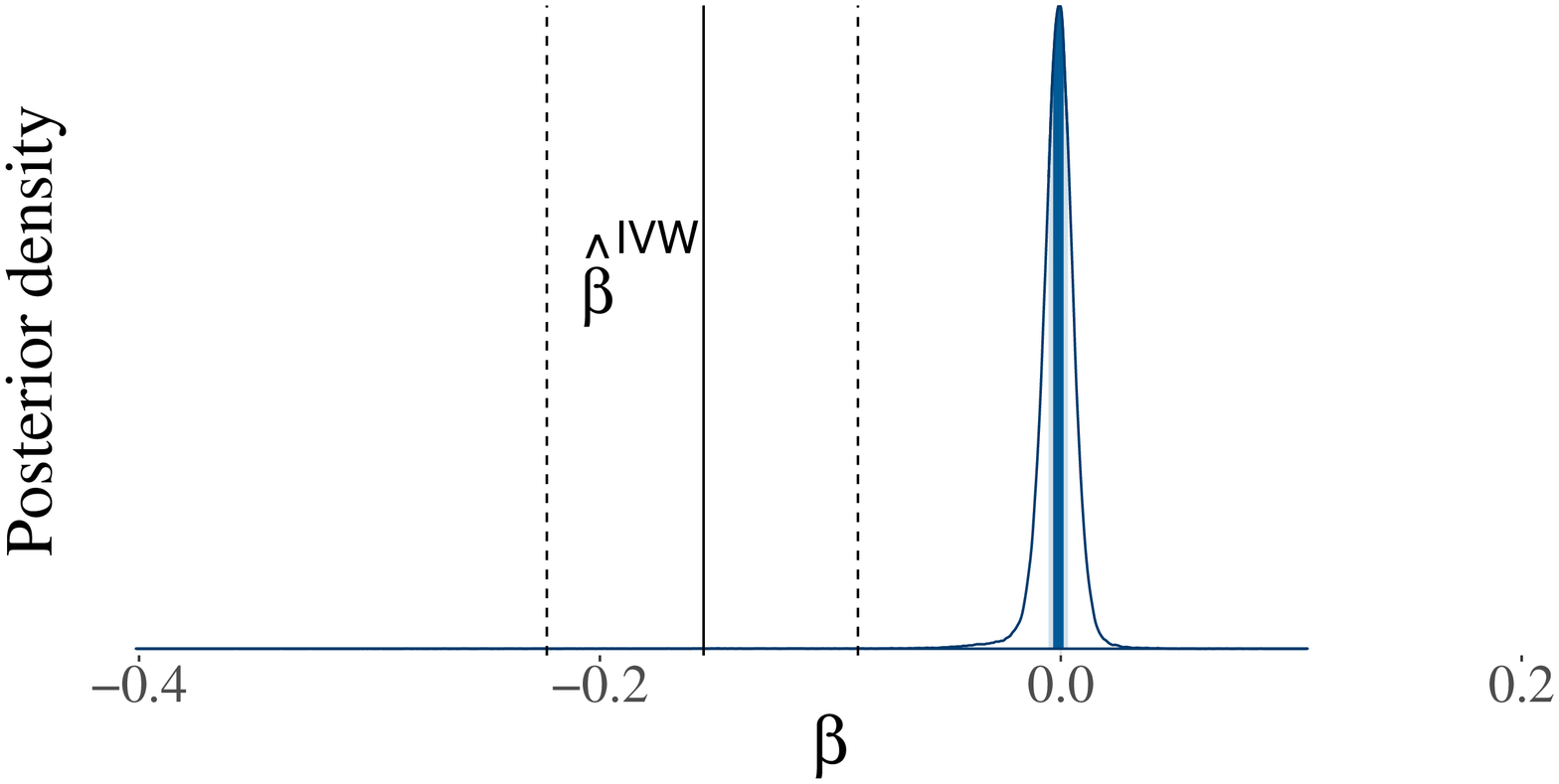}
			\caption{Estimated posterior distribution for the causal effect of birth weight on adult fasting glucose levels. The light shaded area in the posterior represents the interquartile range (IQR), while the dark shaded line indicates the median. For the Gaussian mixture prior in Equation~\eqref{eqn:gauss_mix_prior}, we have taken $\tau^2 = 1$ and $\lambda = 10^{-4}$. The IVW estimate reported in Del Greco et al.~\cite{del_greco_detecting_2015} and its confidence bounds are shown for comparison.} \label{fig:BW_FG_beta} %
		\end{minipage} \hfill
		\begin{minipage}{0.49\linewidth}
			\includegraphics[width=\textwidth]{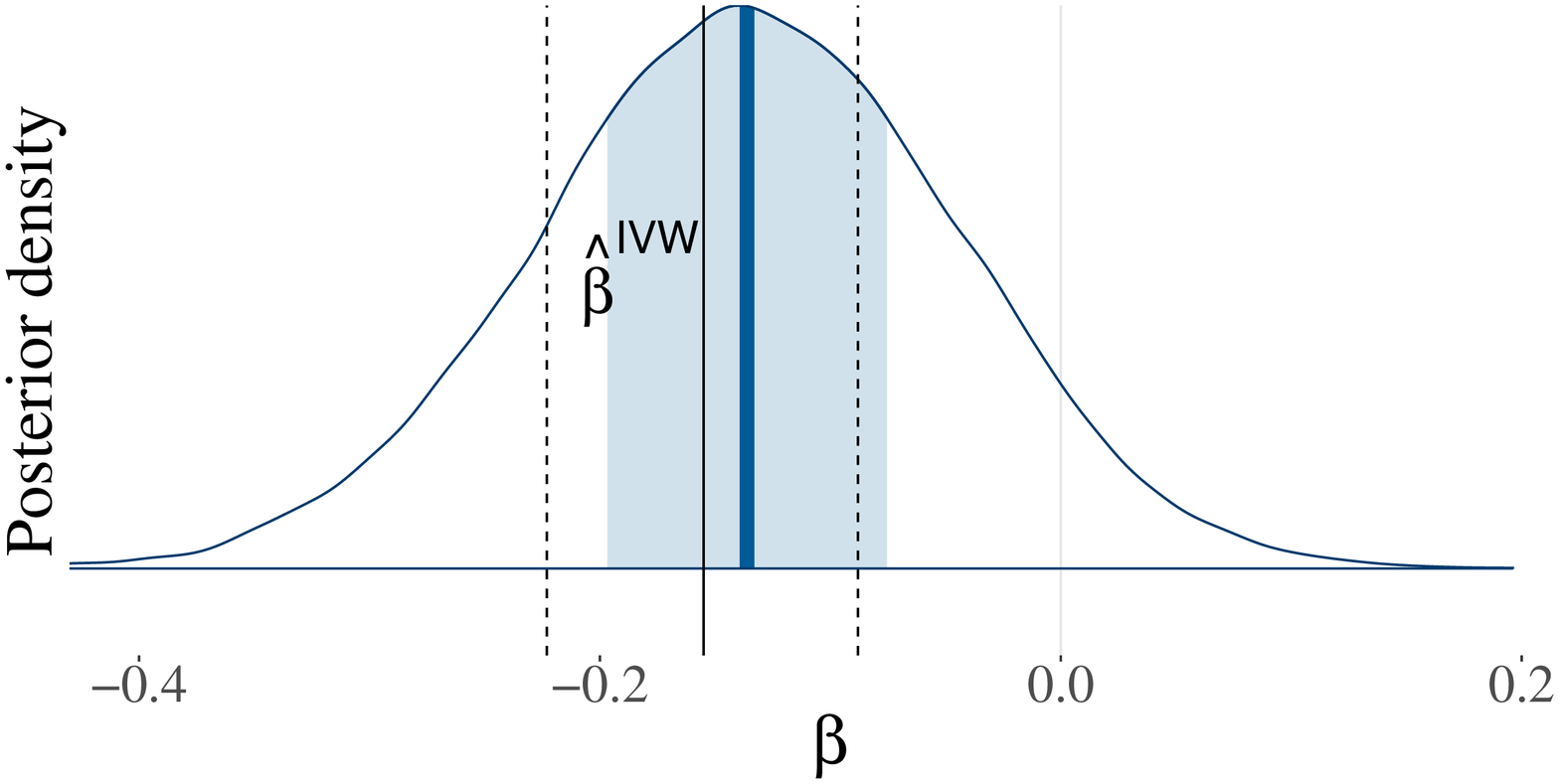}
			\caption{By adapting the prior knowledge to fit the classic instrumental variable setting, we are able to recover the IVW estimate. The light shaded area in the posterior represents the interquartile range (IQR), while the dark shaded line indicates the median. For the Gaussian mixture prior in Equation~\eqref{eqn:gauss_mix_prior}, we have taken $\tau^2 = 1$ and $\lambda = 10^{-4}$, respectively. The IVW estimate reported in Del Greco et al.~\cite{del_greco_detecting_2015} and its confidence bounds are shown for comparison.} \label{fig:BW_FG_beta_adapt}
		\end{minipage}
	\end{figure}
	
	We ran our model using the summary statistics for the genotype-exposure and genotype-outcome associations provided in Del Greco et al.~\cite{del_greco_detecting_2015} (Table IV). The regression coefficient for the observational exposure-outcome association is taken from Daly et al.~\cite{daly_low_2005}, who reported a decrease of 0.01 $mmol/L$ in fasting glucose levels per kilogram increase in birth weight in a retrospective cohort study on a multiethnic sample of 855 New Zealand adolescents. The regression coefficients they computed were adjusted for age, sex, and ethnicity. One potential limitation of using this data is that there may be biases induced by the different adjustments made in the study.~\cite{lawlor_commentary_2016} Another potential issue is the fact that the observational study sample is taken from an underlying population which is different in terms of ethnicity and age.~\cite{thompson_mendelian_2016} Nevertheless, the associations found in the observational study are consistent with those found in subsequent studies.~\cite{norris_size_2011}
	
	Our estimated posterior in Figure~\ref{fig:BW_FG_beta} indicates that no causal effect is necessary to explain the observed data. This conclusion differs significantly from that of Del Greco et al.~\cite{del_greco_detecting_2015} because we are making less stringent assumptions regarding the parameter strengths. However, we can arrive at their conclusion if we adapt our prior knowledge by incorporating stronger prior assumptions. To obtain the posterior distribution in Figure~\ref{fig:BW_FG_beta_adapt}, we assumed that the pleiotropic effects are negligible (by setting $\wt_{\plyv} = 0$, equivalent to a weaker form of \texttt{IV3}) and that the instrument strengths are relevant (by setting $\wt_{\isv} = 1$, equivalent to a weaker form of \texttt{IV1}). Additionally, we set $\wt = 0$ in the Gaussian mixture prior (Equation~\eqref{eqn:gauss_mix_prior}) for $\tilde{\cc}_\exs, \tilde{\cc}_\out$ and $\tilde{\ce}$, so as to not penalize (potentially) relevant confounding and causal effects too strongly. Since this new set of assumptions is much closer to the typical assumptions made in an IV analysis, we were able to recover the IVW estimate. Our results show, however, that if we penalize every strong interaction between variables equally (see Subsection~\ref{ssec:prior}), then the posterior estimate of $\ce$ indicates a negligible causal effect of birth weight on adult fasting glucose.

	\subsection{Effect of body mass index on the risk of Parkinson's disease}
		
	\begin{wrapfigure}[22]{r}{0.35\linewidth}
		\vspace{-3\baselineskip}
		\centering
		\includegraphics[width=\linewidth]{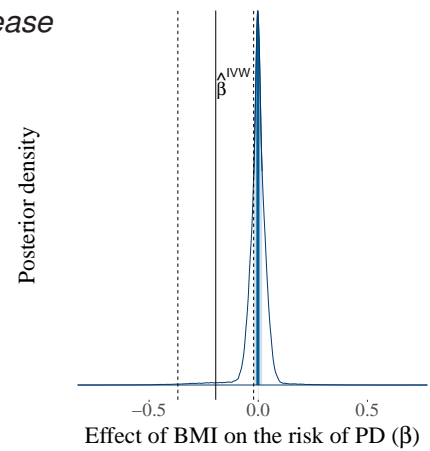}
		\caption{Estimated causal effect of BMI on the risk of PD expressed as the difference in log-odds of PD per 5 $kg/m^2$ increase in BMI. The light shaded area in the posterior represents the interquartile range (IQR), while the dark shaded line indicates the median. The IVW estimate $\hat{\ce}^{IVW}$ derived from~\eqref{eqn:ivw} along with its 95\% confidence bounds are shown for comparison.}
		\label{fig:Pks_BMI_beta}
	\end{wrapfigure}
	
	There have been conflicting findings regarding the association between body mass index (BMI) and Parkinson's disease (PD) in observational studies. Both positive and negative associations between higher BMI and the risk of PD have been reported.~\cite{noyce_estimating_2017} Wang et al.~\cite{wang_body_2015} performed a meta-analysis of ten cohort studies and their obtained pooled risk ratio (RR) for the association of a 5 $kg/m^2$ higher BMI with the risk of PD was 1.00 (95\% CI [0.89, 1.12]), which suggests that BMI is not associated with the risk of PD. In a Mendelian randomization analysis, Noyce et al.~\cite{noyce_estimating_2017} estimated the causal effect of BMI on the risk of PD by combining the ratio estimates corresponding to 77 selected SNPs via an inverse-variance weighted scheme. The results of their analysis suggest a putative protective causal effect of BMI. More specifically, their MR analysis yielded an IVW log-odds ratio (log-OR) of -0.195 (95\% CI [-0.368, -0.021]), meaning that a lifetime exposure of 5 $kg/m^2$ higher BMI was associated with a lower risk of PD.
	
\begin{figure}[!htb]
		\begin{minipage}{0.59\linewidth}
			\centering
			\includegraphics[width=\textwidth]{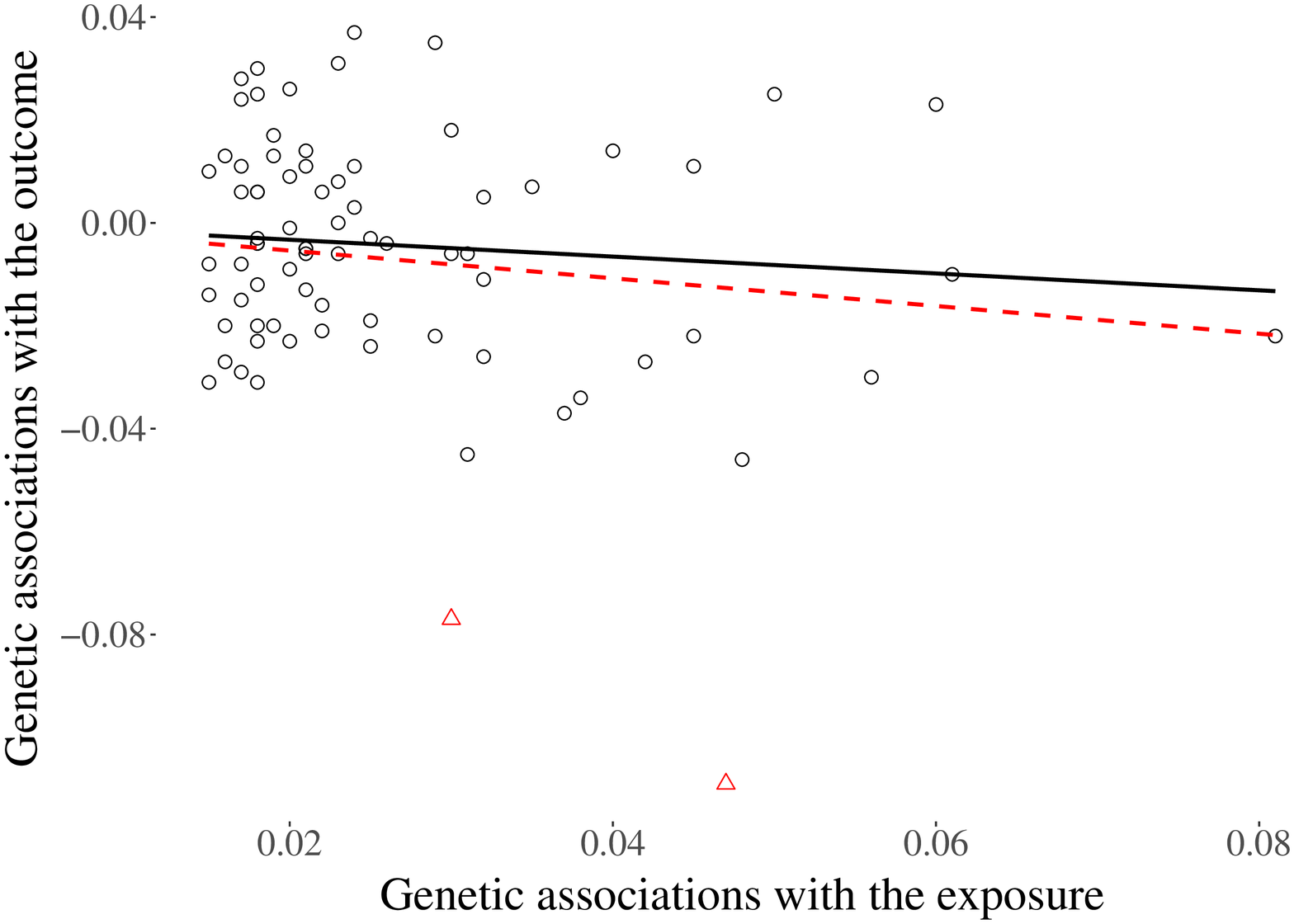}
			\caption{Scatter plot of the genetic associations with BMI (horizontal axis) and PD risk (vertical axis) for 77 genetic variants. The two outliers (triangles) show a relatively strong association with the outcome given their association with the exposure. The regression line including the outliers is dashed, while the regression line obtained without the outliers is continuous.}
			\label{fig:Pks_BMI_outliers}
		\end{minipage} \hfill
		\begin{minipage}{0.39\linewidth}
			\centering
			\includegraphics[width=\linewidth]{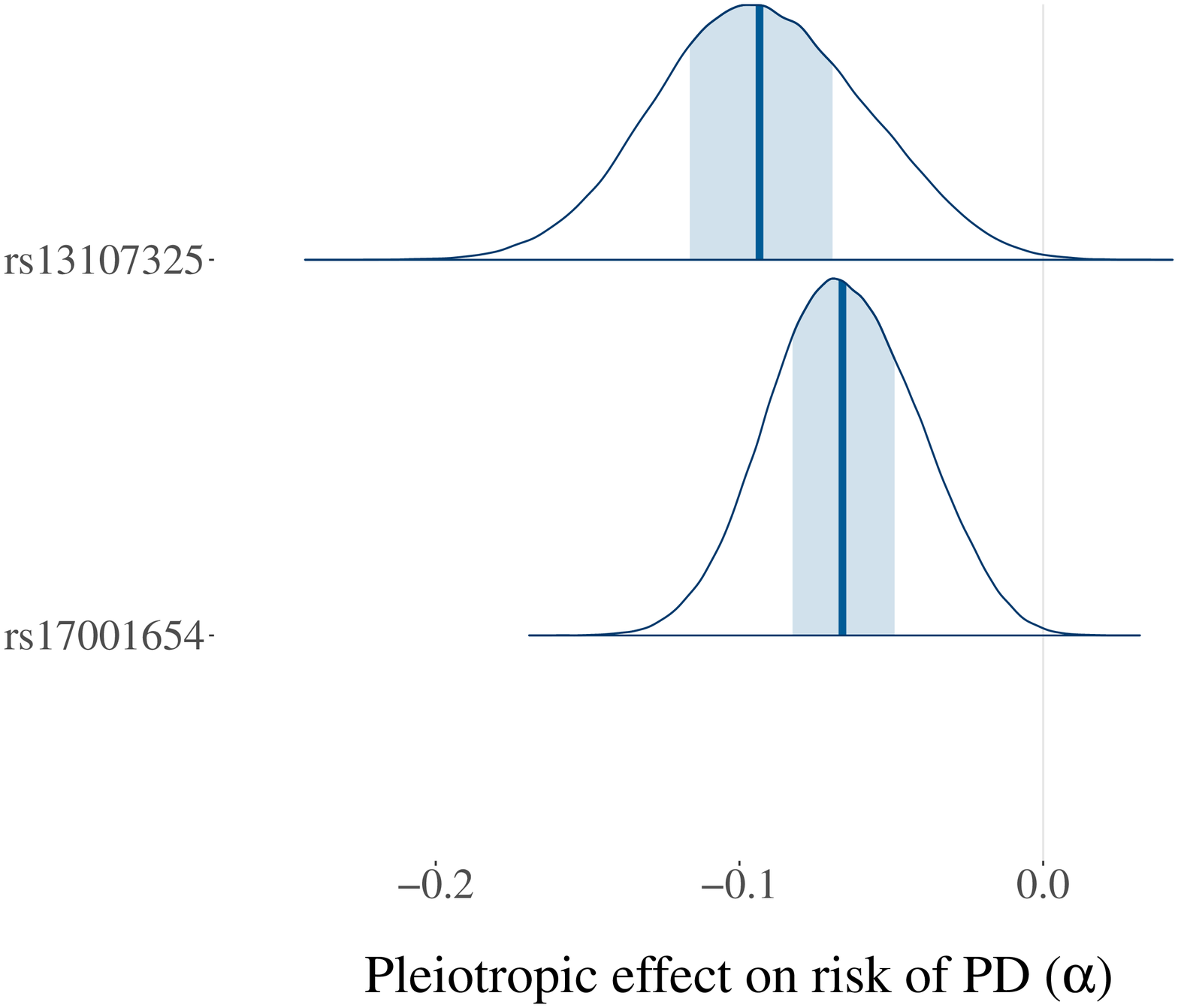}
			\caption{Estimated pleiotropic effects for the two genetic variants suspected of being pleiotropic outliers (\texttt{rs17001654} and \texttt{rs13107325}). The light shaded area in the posterior represents the interquartile range (IQR), while the dark shaded line indicates the median. Most of the posterior mass is distributed away from zero, thereby supporting the suspicion that these two variants exhibit horizontal pleiotropy.}
			\label{fig:Pks_BMI_alpha}
		\end{minipage} 
	\end{figure}

	We also took the log-odds ratio as the continuous outcome variable. We used the summary data from Noyce et al.~\cite{noyce_estimating_2017} for the genetic associations with BMI and the log-OR of PD. We used the RR reported by Wang et al.~\cite{wang_body_2015} as an approximation for the OR, since the incidence of PD is very low, resulting in an estimated log-OR of zero for the observational association of a 5 $kg/m^2$ higher BMI with the risk of PD. We fitted our model assuming the direction of causality from BMI to PD and we obtained a causal effect estimate centered around zero, with 96.3\% of the probability mass in the interval [-0.1, 0.1] (see Figure~\ref{fig:Pks_BMI_beta}). This result casts doubt on the existence of a causal link between BMI and the risk of PD.
	When looking at the scatter plot comparing the genetic associations of the 77 variants with the outcomes and their associations with the exposures (Figure~\ref{fig:Pks_BMI_outliers}), we observe the existence of two outliers, corresponding to the red triangles. These two variants show a low association with the outcome relative to the others given their strength as instruments. It is possible that the unusually large negative association with the risk of PD is due to unobserved pleiotropic effects. This claim is supported by our estimates of the pleiotropic effect $\alpha$ for these two genetic variants, which are shown in Figure~\ref{fig:Pks_BMI_alpha}. 
	
	To further substantiate our claim that these variants represent outliers for the IVW and MR-Egger analysis, we used the radial regression approach of Bowden et al.~\cite{bowden_improving_2017}. By running the radial IVW and MR-Egger regressions using first order weights and specifying a statistical significance threshold of 0.01, we discovered the same two outliers observed visually in the scatter plot. Alternatively, we used the \textit{Mendelian randomization pleiotropy residual sum and outlier} (MR-PRESSO) test~\cite{verbanck_detection_2018} to identify potential pleiotropic outliers. With MR-PRESSO, we detected the two genetic variants highlighted in Figure~\ref{fig:Pks_BMI_outliers} as outliers for a significance threshold of 0.1. The outlier-corrected causal estimate produced by MR-PRESSO remained `barely' significant at the 0.05 level: $\hat{\ce}$ = -0.1669 (95\% CI [-0.3335, -0.0003]).
	
	To sum up, the results of our investigation suggest that the protective causal effect of BMI on the risk of PD discovered by Noyce et al.~\cite{noyce_estimating_2017} could also be due to some of the genetic variants exhibiting horizontal pleiotropy. We suspect two variants in particular to be pleiotropic outliers in the MR analysis, as suggested by our method and other approaches.~\cite{bowden_improving_2017, verbanck_detection_2018}
	
	\subsection{Does coffee consumption influence smoking?}
	
	In this experiment, we investigated the association between coffee consumption and cigarette smoking. It is unclear whether this association is causal. A Mendelian randomization study performed by Treur et al.~\cite{treur_smoking_2016} provided no evidence for causal effects of smoking on caffeine or vice versa. Bjørngaard et al.,~\cite{bjorngaard_heavier_2017} on the other hand, found that higher cigarette consumption causally increases coffee intake. If a causal link between coffee consumption and cigarette smoking exists, its direction is also unclear. Verweij et al.~\cite{verweij_investigating_2018} employed a two-sample bidirectional Mendelian randomization study to investigate, among other things, the causal association between the use of nicotine and caffeine, but found little evidence of a causal relationship in either direction. In another study, Ware et al.~\cite{ware_does_2017} assessed the impact of coffee consumption on the heaviness of smoking, but also obtained inconclusive results: one of their two-sample MR analyses indicated that heavier consumption of caffeine might lead to reduced heaviness of smoking, while in other MR analyses they found no evidence of a causal relationship between coffee consumption and heaviness of smoking. Ware et al.~\cite{ware_does_2017} concluded it is unlikely that coffee consumption has a major causal impact on cigarette smoking and suggested the possibility of reverse causation, i.e., smoking impacting coffee consumption, or confounding as alternative explanations for the observed association.	
	
	We used the summary statistics reported by Ware et al.~\cite{ware_does_2017} to explore this association. The exposure variable is coffee consumption measured in cups per day, while the outcome variable is smoking measured in cigarettes per day. The summary measurements for the gene-exposure association were taken from the European replication sample ($\nobs \le$ 30 062) of the Coffee and Caffeine Genetics Consortium (CCGC) genome-wide association study meta-analysis.~\cite{cornelis_genome-wide_2015} Eight independent coffee related variants meeting the threshold for genome-wide significance in the trans-ethnic GWAS meta-analysis were considered. The associations of coffee-related SNPs with the outcome were obtained from the UK Biobank ($\nobs =$ 8 072). Ware et al.~\cite{ware_does_2017} also computed an observational association of 0.45 additional cigarettes per day for each consumed cup of coffee among the 8 072 current daily smokers who reported consuming coffee.
	
	\begin{figure}[!htb]
		\begin{subfigure}{0.49\linewidth}
			\includegraphics[width=\linewidth]{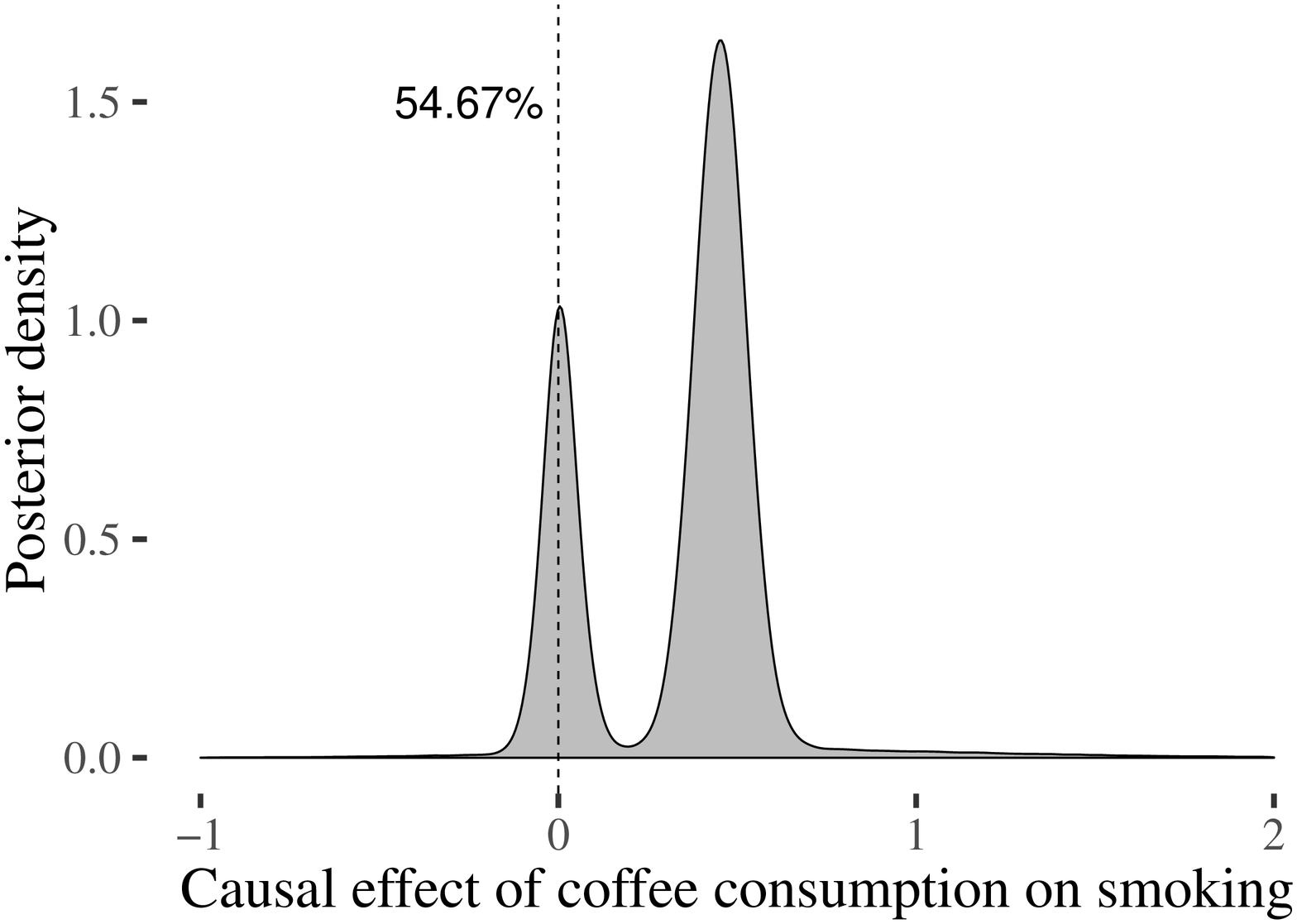}
			\caption{Posterior distribution of the putative causal effect of coffee consumption on smoking. In the case of reverse causation (causal link from smoking to coffee consumption), this effect is zero, as indicated by the vertical dashed line. The estimate next to the line (54.67\%) is the evidence for the reverse model.} \label{fig:posterior_ce_dir}
		\end{subfigure} \hfill
		\begin{subfigure}{0.49\linewidth}
			\includegraphics[width=\linewidth]{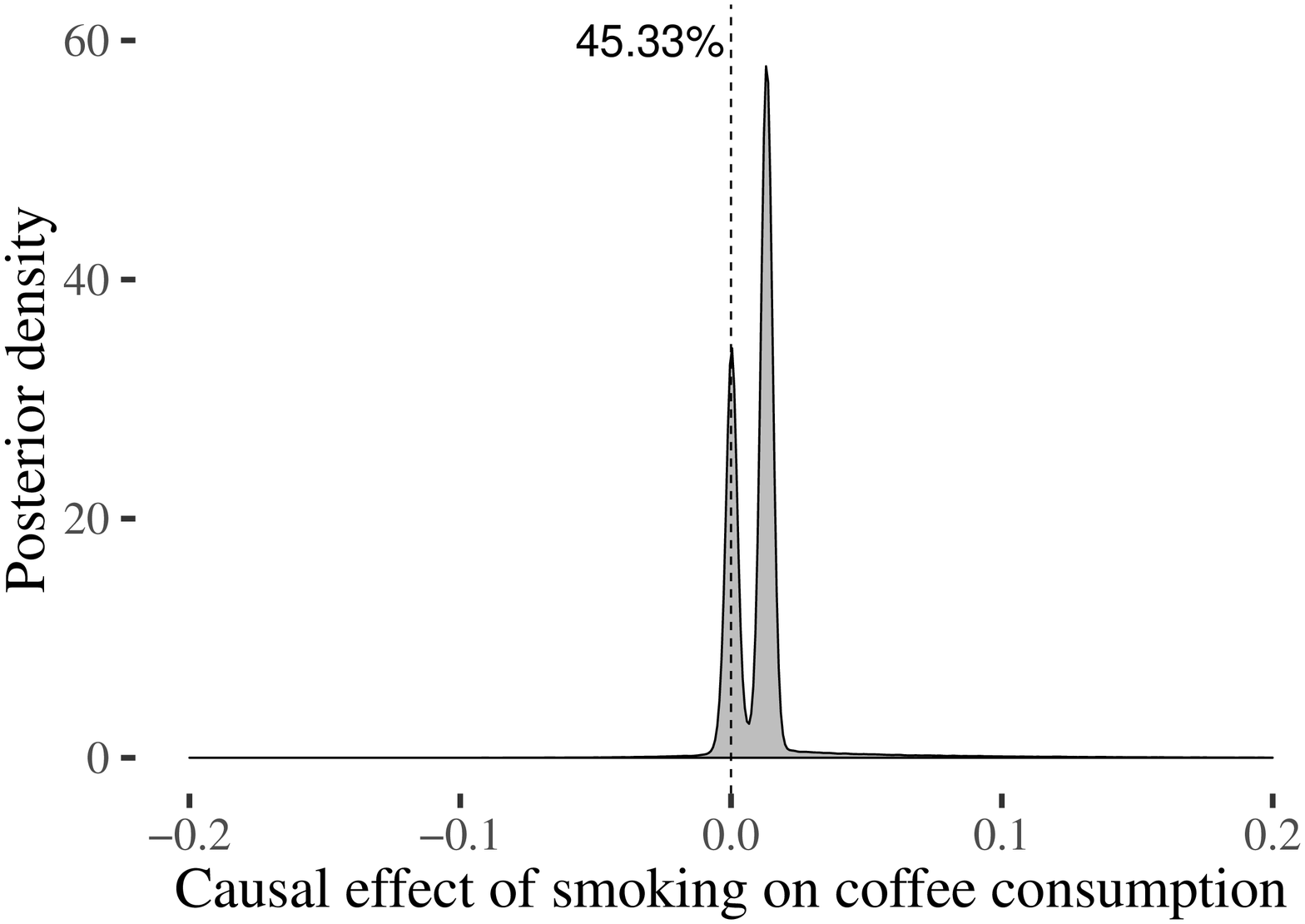}
			\caption{Posterior distribution of the putative causal effect of smoking on coffee consumption. In the case of reverse causation (causal link from coffee consumption to smoking), this effect is zero, as indicated by the vertical dashed line. The estimate next to the line (45.33\%) is the evidence for the reverse model.} \label{fig:posterior_ce_rev}
		\end{subfigure}
		\caption{Comparison of the causal effect estimates between $\exs$ (coffee consumption in cups per day) and $\out$ (heaviness of smoking in cigarettes per day) for the two possible causal directions. The estimated evidence for the two models is $p(\model \given \exs \to \out) = 54.67\%$ and $p(\model \given \out \to \exs) = 45.33\%$, respectively. In the left figure, we see the estimate of $\beta_{\exs \to \out}$, which is the causal effect of coffee consumption on heaviness of smoking, under the assumption that the causal link $\exs \to \out$ exists. In the right figure,  we see the estimate of $\beta_{\out \to \exs}$, which is the causal effect of heaviness of smoking on coffee consumption, under the assumption that the causal link $\out \to \exs$ exists. }
	\end{figure}

	In our approach, we considered two candidate models corresponding to the two possible causal directions. In the first model (Figure~\ref{fig:mr_pleiotropy}), we assumed a causal link from coffee consumption to smoking, while in the second model Figure~\ref{fig:mr_reverse}) we assumed the reverse causal link. We computed the evidence for the two models and then, assuming a-priori that either of the two models is equally likely, we obtained a posterior probability of 42.6\% for the first model and 57.4\% for the second model. The posterior over the causal effect is shown for both directions in Figures~\ref{fig:posterior_ce_dir} and~\ref{fig:posterior_ce_rev}. According to our results, there is significant uncertainty in the direction of the causal effect. This suggests that the observed association may in fact be due to some confounding factor. Still, that does not exclude the possibility of a causal effect between coffee consumption and heaviness of smoking, but more data is required to substantiate this claim.

	\section{Discussion} \label{sec:discussion}
	
	In this paper, we have proposed a Bayesian approach to Mendelian randomization (\textsc{BayesMR}) for finding the probable direction of a causal link between two phenotypes and for estimating the magnitude of its causal effect. The novelty in our method consists in inferring the direction of causality, estimating the causal effects and promoting model sparsity in a single-step approach. This is achieved by using a nested sampling scheme~\cite{skilling_nested_2006, handley_polychord_2015, handley_polychord_2015-1} to compute the model evidence for both causal directions and to sample from the parameter posterior distribution simultaneously.
	
	Many traditional MR methods can be seen as limiting cases of our more general approach. For example, the ``no pleiotropy'' assumption~\cite{bowden_framework_2017} made when using genetic variants as instrumental variables can be incorporated by setting the prior proportion of the slab component for the pleiotropic effect to zero ($\wt_\ply = 0$) and then taking $\lambda \to 0$. Berzuini et al.~\cite{berzuini_bayesian_2018}, for example, have proposed a similar Bayesian approach, but they assume a particular causal direction. This assumption can be incorporated into our model by setting the prior probability of reverse causation to be zero, as shown in Section~\ref{sec:selection}. Our Bayesian solution subsumes standard Mendelian randomization (Figure~\ref{fig:iv_assumptions}) and the LCD pattern (Figure~\ref{fig:lcd}), is robust to the presence of pleiotropic effects (Figure~\ref{fig:mr_pleiotropy}), and incorporates the possibility of reverse causation in the exposure-outcome association (Figure~\ref{fig:mr_reverse}). The resulting output appropriately reflects the uncertainty of having to consider two possible causal directions due to the existence of alternative causal pathways. 
	
	In the era of high-throughput genomics, it is becoming increasingly likely to select invalid instruments as GWAS sample sizes continue to grow.~\cite{hemani_automating_2017} Our approach allows for the selection of a much broader range of genetic variants, since it does not require any of the instruments to be valid. Furthermore, we can select even those variants for which it is not clear whether they primarily influence the exposure or the outcome. In this sense, we envision running \textsc{BayesMR} on a large number of potential genetic candidates collected from genome-wide association studies, which can be related to either the exposure or the outcome. Another advantage of our method is that it does not rely on individual-level data. This is becoming increasingly important in the era of large GWA studies,~\cite{visscher_10_2017} where summary statistics regarding genetic associations obtained from large independent samples are made publicly available. Our approach, however, is currently not designed to handle potentially correlated genetic variants, which means it cannot use variants that are in linkage disequilibrium with each other as input. Another concern is the computational scalability of the Bayesian inference to a large number of instrumental variables. In the future, we plan to explore more scalable nested sampling approaches, for instance the one proposed by Buchner,~\cite{buchner_collaborative_2017-1} and to extend \textsc{BayesMR} to account for correlation among genetic variants.
	
	A key aspect of our approach lies in the structural assumption that interactions between variables are either `weak' (irrelevant) or `strong' (relevant).~\cite{bucur_robust_2017} This assumption is different from the traditional zero-nonzero dichotomy and, in our opinion, more realistic than assuming, for example, that a genetic effect on an outcome variable is ``completely mediated'' by an another variable. The Gaussian mixture prior is a natural choice for expressing this assumption and has the intuitive interpretation of capturing the small, irrelevant effects in the `spike' (lower variance) component and the large, relevant effects in the `slab' (higher variance) component. The mixture Gaussian prior has already been used for example by Li~\cite{li_mendelian_2017} to induce sparsity in the estimation of pleiotropic effects. Here, we extend the usage of this prior to the other structural parameters, which allows for a more general view of Mendelian randomization analyses. For example, putting a Gaussian mixture prior on the strength of instruments will enable us to make an automatic selection of the relevant instruments in a batch of preselected genetic variants, where the proportion of relevant instruments can be determined by learning the shared hyperparameter $\wt_{\isv}$.
	
	Our chosen prior is informative in the sense that it informs which effects we consider a-priori to be relevant. Because of this, care must be taken when setting the prior hyperparameters. While this informativeness may be seen as a weakness of the Bayesian approach, it also empowers the research to input sensible assumptions regarding the expected magnitude of effects in an intuitive fashion. If one has a prior idea regarding which effect sizes are deemed relevant and which irrelevant, then one can appropriately tune the variances of the `spike' and the `slab' to reflect this belief. When it is not clear how to distinguish between relevant and irrelevant effects, one can treat all detectable effects as relevant by letting $\lambda \to 0$ in a first attempt. At the same time, one can choose $\tau$ large enough to give support to effect sizes that are substantively different from zero, but not so large that unrealistic values are supported.~\cite{george_variable_1993} Furthermore, if one has a prior belief regarding the relevance of a particular parameter, this can be expressed in the prior by appropriately setting the $\wt$ parameter, which determines the mixture component proportion. Other sparsifying priors have been proposed for use in MR Bayesian methods, such as the \textit{horseshoe prior}~\cite{berzuini_bayesian_2018} or the \textit{Laplace prior}.~\cite{agakov_sparse_2010} It might be interesting to compare the inferences obtained by using these different priors and to assess how easy it is to incorporate prior biological knowledge for each of them.
	
	One important limitation of the approach proposed here lies in the strong parametric assumptions of linearity and Gaussianity.~\cite{didelez_mendelian_2007} These standard assumptions are commonly made for simplicity and computational convenience in similar works such as those by Thompson et al.~\cite{thompson_mendelian_2017}, Li~\cite{li_mendelian_2017}, or Berzuini et al.~\cite{berzuini_bayesian_2018}, but when they do not hold, the estimands can potentially be far off the mark and their interpretation is rendered incorrect. We have shown that \textsc{BayesMR} is robust against mild violations of linearity and Gaussianity, even though great care must be taken when interpreting the results. In future work, it would be interesting to test our method as well as other established methods against a broader range of violations. The current method is also not directly applicable to discrete phenotypic variables. However, as we have shown in Section~\ref{sec:applications}, we can incorporate log-odds ratios of binary values in our model by applying a logit transformation. With our approach, we allow for violations of the exclusion restriction assumption (\texttt{IV3}) and we even allow for violations of \texttt{IV1} (genetic variant is not robustly associated with the exposure), although this is typically not considered an issue. However, we still rely on the genetic variants being independent from any unmeasured confounding variables (\texttt{IV2}).
	
	For future work, we plan to make our approach even more general by relaxing the \texttt{IV2} assumption, and therefore taking into account the possibility that the genetic variables might be associated with unmeasured confounders. This would also mean that the \texttt{InSIDE} (Instrument Strength is Independent from Direct Effect) assumption, which is required for applying MR-Egger~\cite{bowden_mendelian_2015} or the Bayesian method proposed by Berzuini et al.~\cite{berzuini_bayesian_2018}, does not hold. Another immediate extension to our work is handling potential measurement error for the exposure and outcome variables. Traditional methods such as MR-Egger rely on having no measurement error in the exposure (the so-called \texttt{NOME}~\cite{bowden_improving_2018} assumption), which is difficult to achieve in practical applications and may lead to erroneous results. Finally, we are also interested in analyzing the links between variables that exert a mutual causal influence on each other. For this purpose, we intend to extend \textsc{BayesMR} to handle cyclic causal models.
	
	The code implementing the proposed method will be made publicly available by the authors at \url{https://github.com/igbucur/BayesMR}.
	
%
%

	\bibliography{BayesMR}
	\bibliographystyle{SageV}

%
%
%
%

\end{document}